\documentclass[fleqn,usenatbib]{mnras}
\usepackage{graphicx}
\usepackage{multirow}

\newcommand{\Ms}{$M_{\star}$}

\usepackage{savesym}
\usepackage{amsmath}
\savesymbol{iint}
\usepackage{txfonts}
\usepackage{booktabs}
\usepackage{xcolor}
\restoresymbol{TXF}{iint}
\bibpunct{(}{)}{;}{a}{}{,}
\usepackage{txfonts}
\usepackage[T1]{fontenc}
\usepackage{ae,aecompl}

\usepackage{graphicx}	
\usepackage{amsmath}	

\title[GAMA: Galaxy pairs properties]{Galaxy And Mass Assembly (GAMA):  the interplay between galaxy mass, SFR and heavy element abundance in paired galaxy sets}

\author[Gardu\~no, L. E.]{Gardu\~no, L. E.$^{1}$\thanks{E-mail: luis@inaoep.mx}, Lara-L\'opez, M. A.$^{2,3}$,  L\'opez-Cruz, O.$^{1}$,  Hopkins, A. M. $^{4}$
\newauthor
Owers, M. S.$^{5,6}$, Pimbblet,  K. A.$^{7}$, Holwerda, B. W.$^{8}$,
\\
$^{1}$Instituto Nacional de Astrof\'isica, \'Optica y Electr\'onica (INAOE), Luis Enrique Erro No.1, Tonantzintla, Pue., C.P. 72840, M\'exico\\
$^{2}$DARK, Niels Bohr Institute, University of Copenhagen, Lyngbyvej 2, Copenhagen DK-2100, Denmark\\
$^{3}$Armagh Observatory and Planetarium, College Hill, Armagh BT61 9DG, Northern Ireland, UK\\
$^{4}$Australian Astronomical Optics, Macquarie University, 105 Delhi Rd, North Ryde, NSW 2113, Australia\\
$^{5}$Department of Physics and Astronomy, Macquarie University, NSW, 2109, Australia\\
$^{6}$Astronomy, Astrophysics and Astrophotonics Research Centre, Macquarie University, Sydney, NSW 2109, Australia\\
$^{7}$E.A.Milne Centre for Astrophysics, University of Hull, Cottingham Road, Kingston-upon-Hull, HU6 7RX, UK\\
$^{8}$Department of Physics and Astronomy, 102 National Science Building, University of Louisville, Louisville KY 40292, USA\\
}

\date{Accepted 2020 December 03. Received 2020 November 30; in original form 2020 June 09}

\pubyear{2020}

\begin{document}
\label{firstpage}
\pagerange{\pageref{firstpage}--\pageref{lastpage}}
\maketitle


\begin{abstract}
We study the star formation rate (SFR), stellar mass (\Ms) and the gas metallicity (Z) for 4,636 galaxy pairs using the Galaxy And Mass Assembly (GAMA) survey. Our galaxy pairs lie in a redshift range of 0 $<$ $z$ $<$ 0.35, mass range of 7.5 $<$ log( M$_{\star}$/ M$_\odot$) $<$ 11.5 and $\Delta V$ $<$ 1000 km s$^{-1}$. We explore variations in SFR and Z from three point of views: multiplicity, pair separation and dynamics.
We define multiplicity as the number of galaxies paired with a single galaxy, and analyzed for the first time variations in SFR and Z for both, single pairs and pairs with higher multiplicity. For the latter, we find SFR enhancements from 0.025-0.15 dex, that would shift the M-SFR relation of single pairs by 27$\%$ to higher SFRs. The effect of Z on the other hand, is of only 4$\%$.
We analyze the most and least massive galaxy of major/minor pairs as a function of the pair separation. We define major pairs those with mass ratios of 0.5 $<$ $M_1$/$M_2$ $<$ 2, while pairs with more discrepant mass ratios are classified as minor pairs. We find SFR enhancements of up to 2 and 4 times with respect to their control sample, for major and minor pairs. For the case of Z, we find decrements of up to 0.08 dex for the closest pairs.
When we focus on dynamics, Z enhancements are found for minor pairs with high velocity dispersion $(\sigma_p > 250 \; \mathrm{km\,s
^{-1}})$ and high multiplicity.
\end{abstract}



\section{Introduction}

The evolution of galaxies is influenced by the environment in which they reside (filaments, clusters, groups, pairs or voids) and with the level of interaction they have with their neighbours. Accordingly, groups and pairs of galaxies are excellent laboratories to investigate the effects of galaxy interaction and baryonic material exchange between members. Pairs of galaxies have been studied extensively since they can be considered the minimal structures in the cosmic web. 

It is thought that  galaxy interactions play an important role in their evolution, since it could affect their star formation rate \citep{1996ApJ...464..641M, 2007A&A...468...61D} and Z \citep{2019A&ARv..27....3M}. However, other related properties, could also change, for instance star formation history \citep{2009IAUS..254...67J}, gas fraction \citep{2020A&A...635A.197D}, molecular gas \citep{2018ApJ...868..132P} or morphology \citep{2004ARA&A..42..603K}.

With the advent of large extragalactic surveys \citep[e.g., SDSS][]{2000AJ....120.1579Y}, it is possible to analyze in a robust way the effect of galaxy interactions on their properties. For instance, enhancements of SFR at close projected separations are reported by \citet{Ellison08} even in low density environments \citep[see also][]{2010MNRAS.407.1514E}. \citet{2012MNRAS.426..549S} confirm such SFR enhancements in close projected separations, although they also find evidence of considerable enhancements for larger separations.

The gas metallicity in galaxy pairs was also studied by \citet{Ellison08}, who found decrements of metallicity for close projected separations. \citet{2012MNRAS.426..549S} found even higher decrements in metallicity for the same close separations, but they also report that such decrements are preserved up to intermediate separations. Furthermore, by analyzing merger systems, it was found that the gas in post-merger systems is metal poor \citep{2013MNRAS.435.3627E}. 

Some effects are particularly strong in major or minor pair/mergers depending on the mass ratio of both galaxies \citep{2008MNRAS.384..386C}. Detailed analysis find that galaxies in major mergers tend to have an enhanced star formation, while this is suppressed in the least massive galaxy of minor mergers \citep{2015MNRAS.452..616D}.
Even more, radical enhancements in the SFR are reported for post-merger systems \citep{2013MNRAS.435.3627E} and ultra-luminous infrared galaxies (ULIRGs) that are in merger systems \citep{2020A&A...635A.197D}, in comparison with other stages of interaction.
However, there are some controversies regarding such SFR enhancements, since strong episodes of SFR can occur in galaxies that are not interacting \citep{2007MNRAS.381..494O, 2007A&A...468...61D, 2019A&A...631A..51P}. 

The environment of galaxy pairs could affect the properties of both galaxies. For instance, there is evidence that in environments with an intermediate density, mergers and galaxy interactions are more frequent, with important processes (e.g. triggering a starburst) that regulates some galaxy properties (e.g. morphology) \citep{2009MNRAS.399.1157P,2007MNRAS.381..494O}. The M-SFR and M-Z relations could also change as a function of the environment. \citet{2012MNRAS.423.2690S} find a high enhancement of SFR for isolated groups at large scale environments; however \citet{2018MNRAS.481.3456C} do not find a dependency in the M-SFR relation with the environment. Regarding the M-Z relation, a dependence has been found between metallicity and environment because galaxies in high overdensity regions and in clusters have high metallicity \citep{2008MNRAS.390..245C,2009MNRAS.396.1257E}

An alternative approach to study the effect of environment is to use the dynamic information of the galaxies in the system. This can be difficult to evaluate since we can only have a limited knowledge of the orbits in the system \citep[e.g.,][]{2004dmu..book...71T}. Nevertheless, the formalism and methodology employed by \citet{2019ApJ...886L...2L}, which is based on studies from \citet{2015AJ....149...54T} and \citet{1985ApJS...58...39B}, have shown that a dynamic treatment reveals the real link between galaxies in compact groups and can be applied to any overdensity.

From a theoretical point of view, simulations have confirmed some observational results. \citet{2012MNRAS.426..549S} present a suite of 16 galaxy mergers, varying only the galaxy orientations. When the pair reaches 20 kpc, the SFR starts to increase and peaks (for first time) after the first encounter, then separation is presented up to 60 kpc due to dynamics; after this, both galaxies are undergo a second encounter, which ends with the coalescence of both galaxies and just after this event, the SFR has the most extreme enhancement, up to 3 times higher than the first peak. Similar results, with a merger system, are given by \citet{2013MNRAS.433L..59P} who find the strongest enhancements in SFR at the smallest separations, such enhancements are produced in highly disturbed systems approaching their final coalescence. Talking about metallicities, merger simulations in \citet{2012MNRAS.426..549S} show a behaviour that is opposite to the SFR, i.e. while the SFR increases, the metallicity decreases. They show how at 20 kpc the decrements in metallicity start, and reach their first important decrement after the first encounter. Metallicity is mildly enhanced after the first encounter as the galaxies separate. Then, there is a second encounter and the consequent coalescence, at this point the most extreme decrement in gas metallicity is reached.

\citet{2007A&A...468...61D}, \citet{2012MNRAS.426..549S} and \citet{2015MNRAS.449.3719S} agree with the idea that the first pericenter passage could affect the quantity of gas in the disk of galaxies and thus the first episodes of SF. This means that different samples of galaxy pairs must be explored in detail in order to determine which stage of interaction they are in.

Most of the works cited in this section include galaxies that have been paired with more than one galaxy, a concept known as multiplicity, and described in detail in \S\ \ref{PairSelection} of this paper. Galaxy pairs in large surveys such as SDSS report only $\sim$5$\%$ of galaxy pairs at high multiplicities \citep{Ellison08}. Nonetheless, the effect of high multiplicity in galaxy pairs so far has been neglected.

In this paper, we analyze simultaneously the SFR and gas metallicity variations of galaxy pairs in the Galaxy And Mass Assembly (GAMA) survey \citep{2011MNRAS.413..971D, 2015MNRAS.452.2087L}, which is two orders of magnitude deeper than SDSS.  We take advantage of the high close pair completeness of the GAMA survey, which is greater than 95$\%$ for all galaxies with up to five neighbours within 40 arcsec \citep[see][]{2011MNRAS.413..971D, 2011MNRAS.416.2640R} that allow us to study galaxy pairs at high multiplicity. Moreover, deeper surveys with higher completeness such as GAMA report $\sim$20$\%$ of galaxy pairs at high multiplicity (see \S\ \ref{SampleSelection} of this paper). The effect of galaxy pairs at high multiplicity is analyzed in this paper for the first time.

Additionally, since pair separation of galaxy pairs could be biased by projection effects, we propose an analysis of their dynamic nature. We suggest that the velocity difference or velocity separation can be considered as a complementary parameter to the projected pair separation. Thanks to the higher completeness of the GAMA survey, we are able to explore the effects on SFR and metallicity as a function of velocity separation.

The general structure of this work is described as follows. In \S 2 we present a brief review of the GAMA survey as well as the data used. In \S 3 we analyze variations in SFR and metallicity of galaxy pairs as a function of their multiplicity. In \S 4 we study the effects of pair separation on the SFR and Z of galaxy pairs. In \S 5 we investigate how the effect of dynamics change the SFR and Z for the pairs. Finally in \S 6 and \S 7 we present a general discussion, and a summary of our main conclusions, respectively. The cosmological values that we adopt for this study are $\rm{\Omega_\Lambda}$ = 0.7, $\rm{\Omega_M}$ = 0.3, h = 0.7 and $\rm{H_0}$ = 70 km $\rm{s^{-1}}$ $\rm{Mpc^{-1}}$.


\section{Sample selection}\label{SampleSelection}

We use data from the Galaxy and Mass Assembly (GAMA) survey \citep[][\citealp{2015MNRAS.452.2087L}]{2011MNRAS.413..971D}. GAMA is a spectroscopic survey carried out at the Anglo Australian Observatory (AAO) with the 3.9 m Anglo-Australian Telescope (ATT), using the 2dF fibre feed and the AAOmega multi-object spectrograph \citep{2012SPIE.8446E..54S}. Spectra were obtained with 2 arcsec diameter fibers with a spectral coverage from 3700 to 8300 {\AA}, and spectral resolution of 3.2 {\AA} \citep{2013MNRAS.430.2047H}. GAMA collected spectroscopic data of $\sim$300,000 galaxies, covering 280 $\rm{deg}^2$ of the sky with a Petrosian magnitude limit of $r_{pet}<19.8$ mag \citep{2015MNRAS.452.2087L}. GAMA has surveyed a total of $\sim$286 deg$^2$ split into five independent regions; three equatorial called G09, G12, and G15 of 12$\times$5 deg$^2$ each, and two southern  fields, G02 and  G23, of 8.6$\times$6.5 deg$^2$ and 13.8$\times$5 deg$^2$, respectively. We refer to data from the equatorial regions as GAMA-1, and the full five regions as GAMA-II.

\subsection{Pair selection}\label{PairSelection}

We use the GAMA Galaxy Group Catalogue (G$^3$C) described in \citet{2011MNRAS.416.2640R}. The G$^3$C provides catalogues of groups and pairs created under the Friends of Friends (FoF) algorithm based on galaxy-galaxy links as scale of association. \citet{2011MNRAS.416.2640R} define close pair as: 

\begin{eqnarray*}
    P_{r20v500}&=&\{r_{sep}<20\,\mathrm{h^{-1}\,kpc} \wedge v_{sep}<500\; \mathrm{km\,s^{-1}}\},\\
    P_{r50v500}&=&\{r_{sep}<50\,\mathrm{h^{-1}\,kpc} \wedge v_{sep}<500\; \mathrm{km\,s^{-1}}\},\\
    P_{r100v1000}&=&\{r_{sep}<100\,\mathrm{h^{-1}\,kpc} \wedge v_{sep}<1000\; \mathrm{km\,s^{-1}}\}
    \end{eqnarray*} 

Where $r_{sep}$ is the spatial separation of the pair and $v_{sep}$ is the radial velocity separation.

Under the above definition, a single galaxy can be paired with more than one galaxy, therefore in this work we define multiplicity as the number of galaxies that make a pair with a certain galaxy. For instance, a galaxy paired with only one galaxy is classified as multiplicity 2 (M=2). A  galaxy that is paired with another two galaxies is defined as having multiplicity 3 (M=3) and so on. It is important to note that our definition of multiplicity refers to each individual galaxy. For instance, if galaxy A  has been paired with galaxies B and C, then galaxy A has multiplicity 3. However, it does not necessarily mean that galaxies B and C are paired, it will depend on the geometry of the system whether or not the pair definition is fulfill. Following the same example, if the three galaxies were forming a line with galaxy A in the middle, it is likely that galaxy B would be paired only with galaxy A, and thus have multiplicity 2, similarly, galaxy C would be paired with galaxy A and have multiplicity 2 as well.

Our final pair sample has 4,636 galaxies with stellar masses ranging from 7.5 $<$ log( M$_{\star}$/ M$_\odot$) $<$ 11.5, relative velocities $\Delta V$ $<$ 1000 km s$^{-1}$ and separations up to 100 ${\rm h}^{-1}$ kpc. All the galaxies in this work lie in a redshift range of 0 $<$ $z$ $<$ 0.35. From all our galaxy pairs sample, we have 1100 of galaxies (23.7$\%$) at multiplicities higher than 2. Previous works using SDSS report only $\sim$5$\%$ of galaxy pairs at high multiplicities \citep{Ellison08}. We attribute this difference to the higher completeness of the GAMA survey. Since GAMA is two orders of magnitude deeper than SDSS, fainter galaxies are observed and cataloged.

\begin{figure}
\center
\includegraphics[trim={0cm 3cm 0.5cm 3.3cm},clip,scale=0.48]{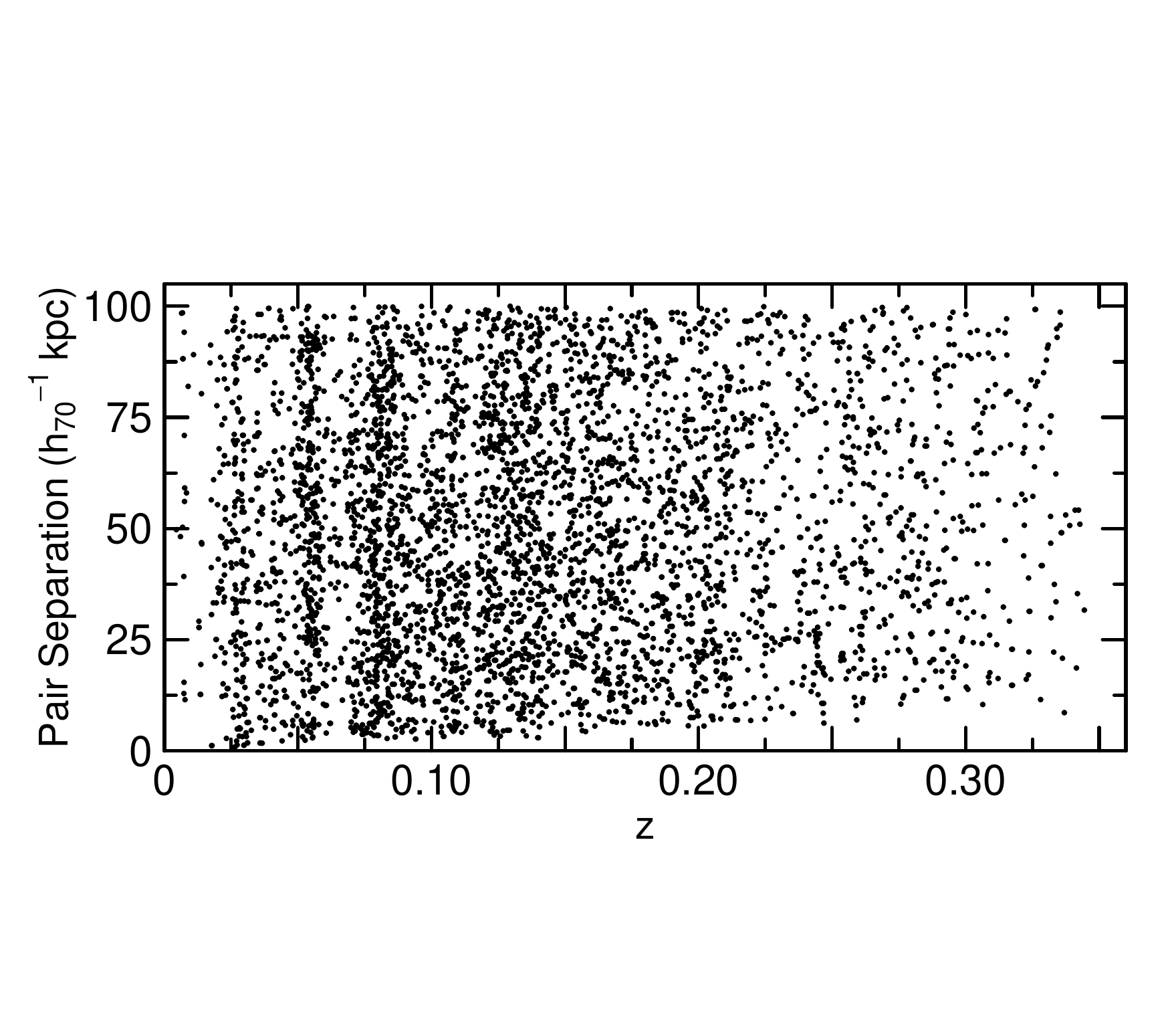}
\caption{Redshift vs. pair separation for galaxy pairs.}
\label{psz}
\end{figure}

In Fig.\ref{psz} we show the redshift vs. pairs separation for our galaxy pair sample. As can be seen, the sample decreases when z $>$ 0.2. Thus, we decide to analyze our sample in three redshift ranges to control for incompleteness. First, the whole sample from 0 $<$ z $<$ 0.35 to be able to compare directly with \citet{Ellison08} in the same redshfit range. Next, from 0 $<$ z $<$ 0.1 to consider when our sample is most complete, and finally from 0.1 $<$ z $<$ 0.2, which is the redshift limit in which our sample start showing signs of incompleteness. Additionally, to further explore any effect of incompleteness, we created a volume limited sample constructed by setting limits in redshift (0-0.1) and Petrosian $r$-band absolute magnitude (M$_r$ $<$ -18.36), see  Fig. \ref{GAMAz01box}.

\subsection{Stellar mass, star formation rate and metallicity estimation}

The data use in this paper are taken from the SpecLineSFRv05 catalogue with the equivalent widths and line flux measurements for GAMA-II spectra. The fitting is made in 5 spectral regions that contain 12 important emission lines [O${\rm{II}}$]$\lambda$3726, $\lambda$3729, H$\beta$, [O${\rm{III}}$]$\lambda$4959, $\lambda$5007, [O${\rm{I}}$]$\lambda$6300, $\lambda$6364, H$\alpha$, [N${\rm{II}}$]$\lambda$6548, $\lambda$6583, [S${\rm{II}}$]$\lambda$6716, $\lambda$6731. For a more detailed description see \citet{2017MNRAS.465.2671G}.

\begin{figure}
\includegraphics[scale=0.4]{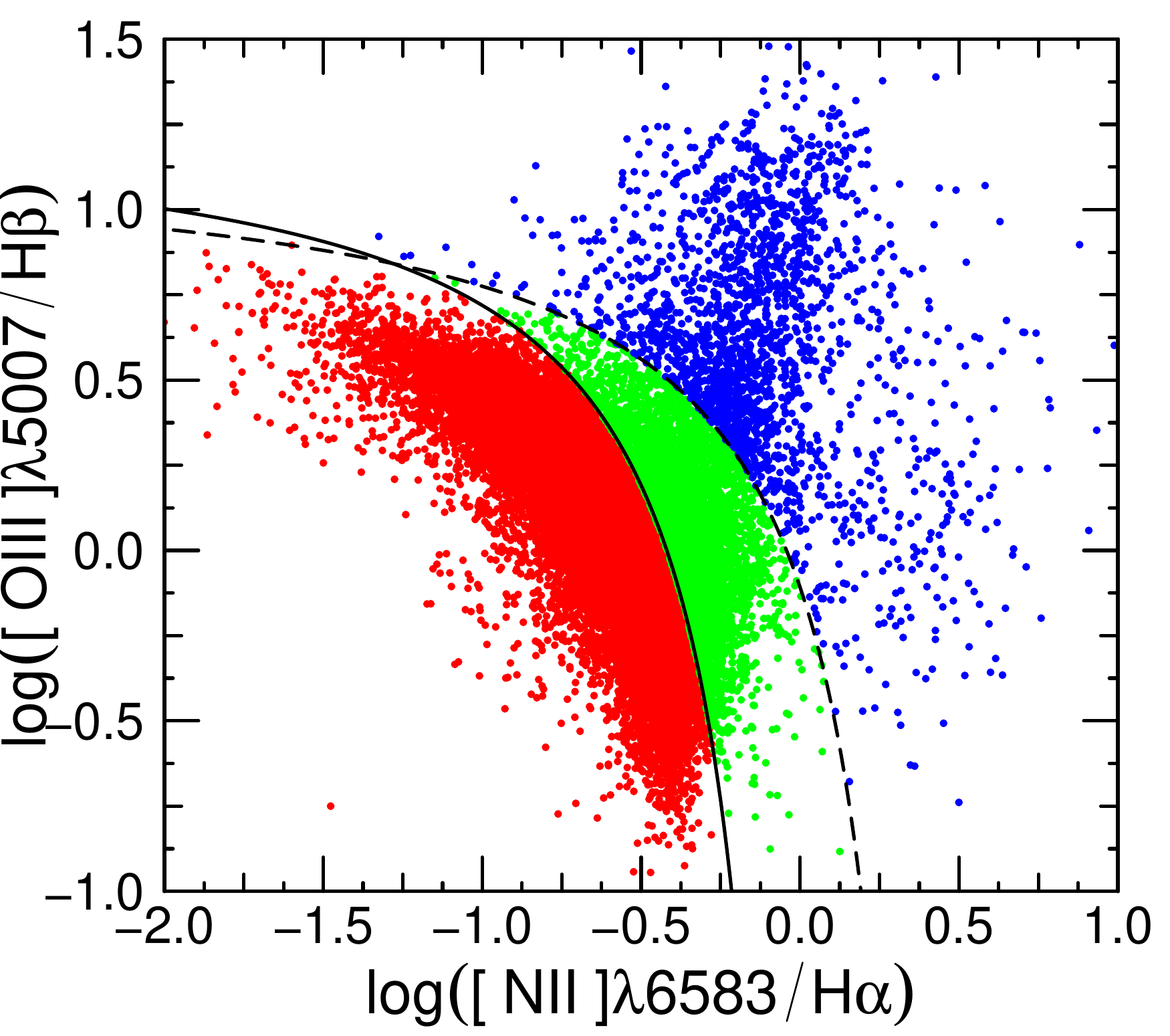}
\caption{BPT diagram for galaxies with  a S/N $>$3 in the four emission lines used. The solid line correspond to the \citet{2003MNRAS.346.1055K} limit, the  dashed line to the \citet{2001ApJ...556..121K} limit. Red, green, and blue colors correspond to  SF, composite, AGN galaxies, respectively.}
\label{BPTSNGAMA}
\end{figure}

\begin{figure}
\includegraphics[scale=0.4]{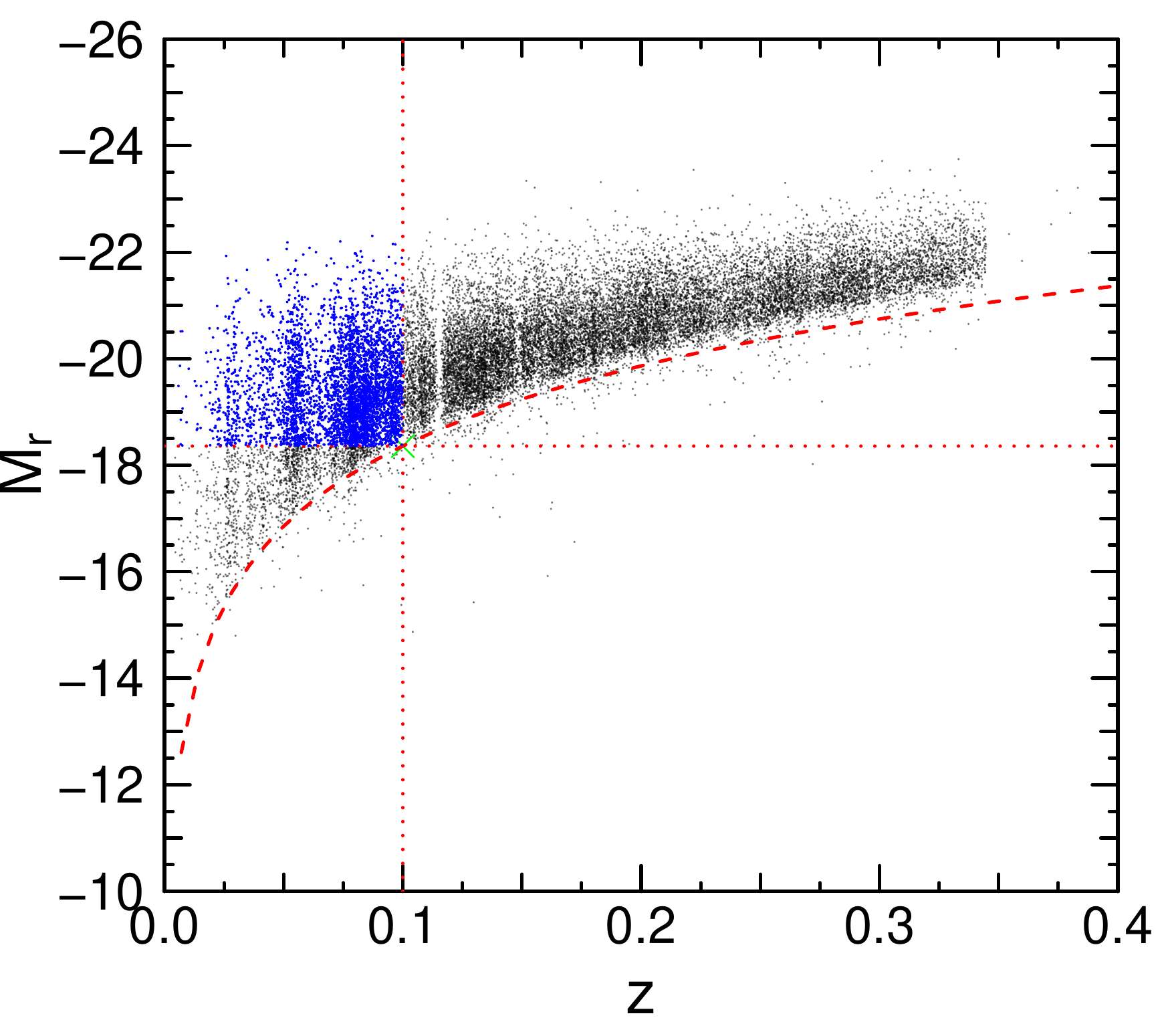}
\caption{\bf Redshift vs. Petrosian $r$-band absolute magnitude (M$_r$) for the main SF GAMA sample (black dots). The blue dots represent a volume limited sample of galaxies that is used in \S \ref{multsec}. The vertical and horizontal dotted red lines are the thresholds for redshift z=0.1, and the absolute magnitude M$_r$=-18.36, respectively. The dashed red line is the distance modulus of the GAMA apparent magnitude limit m$_r$=19.8.}
\label{GAMAz01box}
\end{figure}

For the estimation of SFRs and metallicities, we used the spectroscopic catalogue that contains $\sim$427,000 galaxies. From this emission line catalogue, a cut of S/N$>$3 was applied for the H$\alpha$, H$\beta$, [O${\rm{III}}$]$\lambda$5007 and [N${\rm{II}}$]$\lambda$6583 emission lines, giving us a total of 30,700 galaxies. Next, we selected SF galaxies using the BPT diagram \citep{1981PASP...93....5B} and the classification of \citet{2003MNRAS.346.1055K}. We obtain a total of 24,279 (79\%) SF, 4,449 (15\%) composite and 1,973 (6\%) AGN galaxies, as shown in Fig. \ref{BPTSNGAMA}. Hereafter we call the 24,279 SF galaxies as "the main SF GAMA sample".

SFRs measurements are computed according to \citet{2013MNRAS.433.2764G} using the luminosity of L$_{\rm H\alpha}$, correcting by aperture, obscuration and stellar absorption as follows.

\begin{equation}
\begin{array}{c}\begin{array}{c}\rm{L}_{\rm{H\alpha},\rm{int}}=(\rm{EW}_{\rm{H\alpha}}+EW_c)\times10^{-0.4(M_r-34.10)}\\\end{array}\\\\\times\frac{3\times10^{18}}{\left[6564.61(1+z)\right]^2}\left(\frac{\rm{F_{H\alpha}}/\rm{F_{H\beta}}}{2.86}\right)^{2.36}\end{array}
\label{lumhaeq}
\end{equation}

EW$\rm _c$ is a constant correction for stellar absorption of 2.5\r{A}. M$_r$ is the absolute r-band Petrosian magnitude, $k$-corrected, and corrected by galactic extinction. The last term is the Balmer decrement corrected by stellar absorption via:

\begin{equation}
\frac{\rm{F_{H\alpha}}}{\rm{F_{H\beta}}}=\frac{{\displaystyle\frac{\left(\rm{H\alpha}\;EW+EW_c\right)}{\rm{H\alpha}\;EW}}\times \rm{f_{H\alpha}}}{{\displaystyle\frac{\left(\rm{H\beta}\;EW+EW_c\right)}{\rm{H\beta}\;EW}}\times \rm{f_{H\beta}}}
\end{equation}

f$_{\rm H\alpha}$ and f$_{\rm H\beta}$ correspond to the observed fluxes. The theoretical value 2.86 corresponds to the Balmer Decrement for Case B recombination at an electron temperature of 10,000 K and electron density of 100 cm$^{-2}$ \citep{1989NYASA.571...99O}. The exponent 2.36 in equation \ref{lumhaeq} is determined using the \citet{1989ApJ...345..245C}  dust extinction curve.

SFRs were estimated using the conventional calibration determined by \citet{1989ApJ...344..685K}, with a Salpeter IMF:

\begin{equation}
\rm{SFR}\left[M_\odot\;yr^{-1}\right]=\frac{\displaystyle L_{H\alpha,int}}{1.27\times10^{34}\;W} \times \frac{1}{1.7}
\end{equation}

SFRs are divided by a factor of 1.7 in order to re-calibrate them to a Chabrier IMF, to be consistent with the IMF of the stellar masses (see below).

Metallicity estimates were calculated as in \citet{2013MNRAS.434..451L}. The emission lines were corrected by dust extinction using the reddening coefficient C$_{\rm H\beta}$ and the \citet{1989ApJ...345..245C} extinction curve. We used the O3N2 index and the prescription of \citet{2004MNRAS.348L..59P}: 

\begin{equation}
\rm
O3N2\equiv\log\left[\frac{\left(\lbrack O_{III}\rbrack\lambda5007/H\beta\right)}{\left(\lbrack N_{II}\rbrack\lambda6583/H\alpha\right)}\right]
\end{equation}

\begin{equation}
\rm
12\;+\;\log\;(O/H)_{PP04}=8.73\;-0.32\times O3N2
\end{equation}

In a similar way to \citet{2013MNRAS.434..451L}, we apply a correction to the \citet{2004ApJ...613..898T} calibration:

\begin{equation}
\centering
\rm
{\left[12\;+\;\log\;(O/H)\right]}_{T04}=0.1026\;+1.0211\times{\left[12\;+\;\log\;(O/H)\right]}_{PP04}
\end{equation}

The stellar masses are taken from \citet{2011MNRAS.418.1587T}, who used photometric analysis in the $ugrizYJHK$ bands. For accurate estimations, an error in stellar masses lower than 0.3 dex is required.

In order to test the reliability of our estimations, we compare the M-SFR fit of \citet{2012ApJ...757...54Z} (taken from \citealp{2014ApJS..214...15S}) and the M-Z fit of \citet{2004ApJ...613..898T} with our SF main sample (Fig. \ref{GAMAfid}).

It is noteworthy that the 2 arcsec diameter fibre used by 2dF introduces a bias in the estimated values of the SFRs and metallicity. Nevertheless,
recent studies comparing integrated IFU SFRs from MANGA with its counterparts fibre corrected SFRs from SDSS-DR7, have shown a mean difference of $\sim$0.03 dex \citep[e.g.,][]{Lara19, Ellison2018, Duarte17,2016MNRAS.455.2826R}. The same data show a difference of $\sim$0.03 dex in gas metallicity, although with the fibre spectra biased towards higher metallicities \citep[][]{Lara19}. Since our current analysis is based on differences between pairs and control galaxies selected with the same stellar mass and redshift, we expect a negligible effect due to the 2dF fibre size.


\section{The effect of pair multiplicity on the M-SFR and M-Z relations}\label{multsec}

We follow a similar methodology to \citet{Ellison08}, and create control samples for each galaxy pair for a direct and reliable comparison. To generate control samples, first we created a field-galaxies catalogue by removing galaxies classified as pairs and groups from our main SF GAMA sample, resulting in $\sim$15,521 field galaxies. It is worth noting that our field galaxies catalog might include isolated galaxies, however, we are not imposing any condition to specifically select isolated galaxies. 
Next, using as input the G$^3$C catalogue, we created control samples through an iterative process finding matches in redshift and stellar mass of paired galaxies from the field-galaxies catalogue. The iteration process finishes when the redshift and stellar mass distribution of the control sample matches the pairs sample (we establish an error less than $\pm$0.03 dex). 

The above procedure to create control samples is used for individual galaxy pair sub-samples after we apply cuts in multiplicity and pair separation, as detailed in the following sections. The key idea is to compare galaxy pairs samples with their respective control samples to find differences in SFR and metallicity. The final numbers are shown in Table \ref{samples}. 
\begin{table*}
\centering
\begin{tabular}{cccccccc}
\toprule
\toprule
 &  & \multicolumn{6}{c}{Number of galaxies} \\
\cmidrule{3-8}
 &  & \multicolumn{2}{c}{0 $<$ z $<$ 0.35} & \multicolumn{2}{c}{0 $<$ z $<$ 0.1} & \multicolumn{2}{c}{0.1 $<$ z $<$ 0.2} \\
\cmidrule{3-8}
 &  & Pairs & Control & Pairs & Control & Pairs & Control \\
\midrule
M=2 &  & 3539 & 13727 & 1156(851) & 3219(2160) & 1552 & 6079 \\
M=3 &  & 805 & 6436 & 356(271) & 2152(1309) & 346 & 3124 \\
M=4 &  & 205 & 1956 & 120(89) & 955(558) & 74 & 886 \\
M$\geq$5 &  & 90 & 899 & 65(43) & 552(226) & 25 & 347 \\
Major (all) &  & 693/681 & 8421/8089 & 193/215 & 2099/2282 & 498/315 & 4066/4007 \\
Minor (all) &  & 1285/1980 & 14012/22081 & 525/762 & 4875/6860 & 518/856 & 6077/10498 \\
Major M=2 &  & 622/505 & 7676/6102 & 159/142 & 1731/1502 & 281/239 & 3772/3084 \\
Minor M=2 &  & 1071/1341 & 11966/15957 & 408/447 & 3892/4307 & 441/591 & 5279/7646 \\
Major M$\geq$3 &  & 71/176 & 788/2126 & 36/73 & 387/815 & 27/76 & 318/1008 \\
Minor M$\geq$3 &  & 211/639 & 2311/7023 & 117/315 & 1175/2909 & 77/265 & 864/3318 \\
\bottomrule
\end{tabular}
\caption{Number of galaxy pairs and their respective control sample. The sub-samples are made by multiplicity (M), mass ratio (major or minor pairs), and redshift. For major/minor pairs is indicated the number of most massive galaxies (left side) and  least massive galaxies (right side). The numbers in parenthesis for the column 0 $<$ z $<$ 0.1 correspond to the volume limited sample.}
\label{samples}
\end{table*}

\begin{figure}
\includegraphics[trim={0cm 3cm 0cm 5.5cm},clip,scale=0.47]{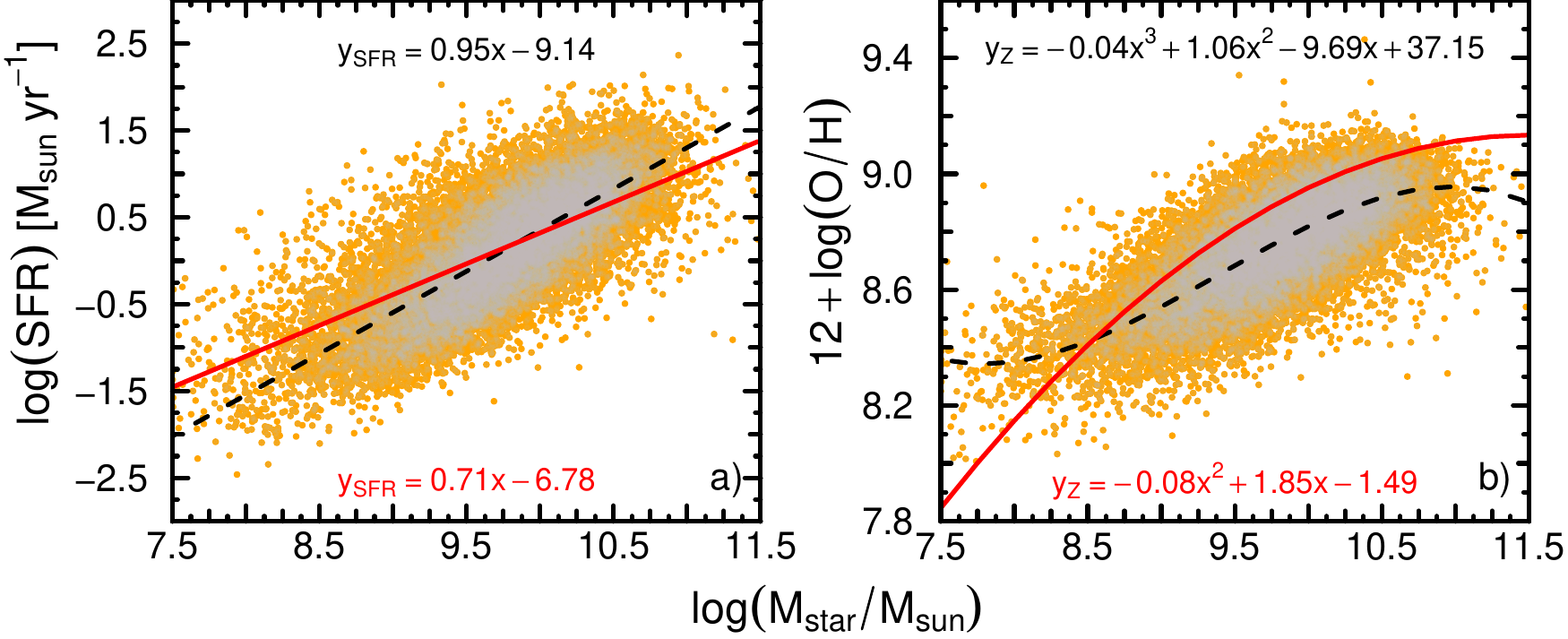}
\caption{The panels show the M-SFR and M-Z relations for our main SF GAMA sample (orange dots), and a comparison with the polynomial fits of \citet{2012ApJ...757...54Z} and \citet{2004ApJ...613..898T} in solid red lines (and bottom equations), respectively. The figure also show the galaxies from our control samples overlapped in grey dots, and their corresponding polynomial fit (fiducial fit) in dashed lines. The corresponding equation of our fits are listed in the top of each panel.}
\label{GAMAfid}
\end{figure}

Since we are analysing sub-samples in redshift, this naturally affects the stellar mass range due to the Malmquist bias, and consequently the shape of the scaling relationships of each sub-sample. To control for this effect in our comparison, first we create a fiducial fit using all galaxies from the control samples. Our fiducial fits for the M-Z and M-SFR relations are shown in Fig. \ref{GAMAfid}.
Next, we fit the M-SFR relation for each control sub-sample by keeping the slope of the fiducial, and fitting only the zero point. The same process is done for each pair sub-sample. Thus, we define the offset $ \rm \Delta SFR=log(SFR)_{pair}-log(SFR)_{control}$, as the difference between the zero points of both linear regressions. Note that $\Delta \rm{SFR}$>0 represents an enhancement of SFR in galaxy pairs and $\Delta \rm{SFR}$<0 represents a decrement, both with respect to their control sample. We compute the residual standard error of the linear regression as $S=\sqrt{\Sigma{(y- y')}^2/n-1}$.

\begin{figure*}
\includegraphics[trim={0cm 7cm 0cm 0cm},clip,scale=0.48]{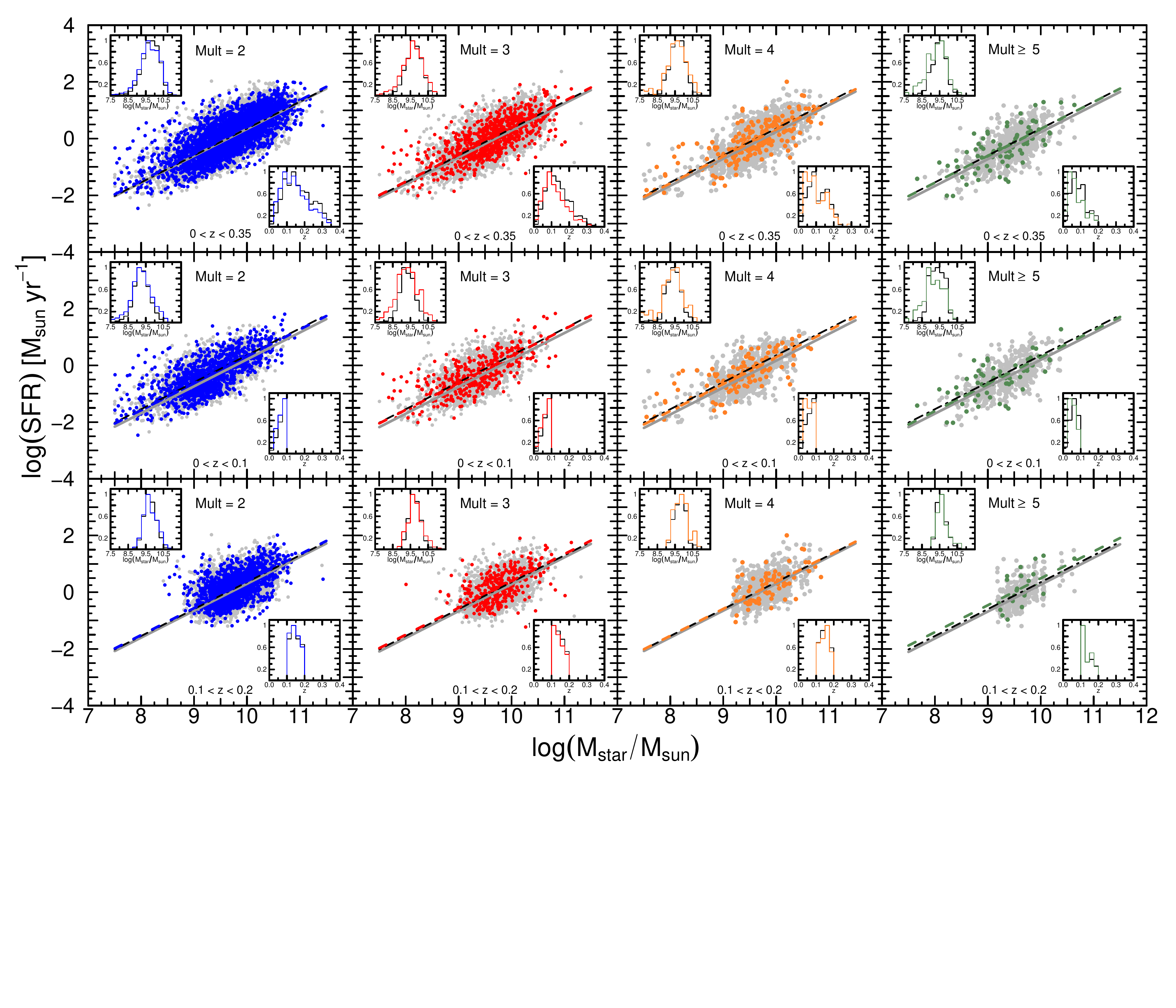}
\caption{The M-SFR relation for different sub-samples of galaxy pairs. The black dashed line represents the fiducial fit to all control galaxies. The gray solid lines represent the fit to the control samples. All the dashed color lines are the fit to the galaxy pairs. Panels from left to right show pair samples by multiplicity, while rows correspond to the different redshift ranges. The gray dots correspond to control galaxies. Colors blue, red, yellow and green correspond to M=2 to M$\geq$5, respectively. The inset histograms in each panel correspond to the comparison between stellar mass (top left corner) and redshift (bottom right corner) between the control and pair samples.}
\label{msfrall}
\end{figure*}

We apply this procedure for M=2 to M$\geq$5, and the following redshift ranges 0 $<$ z $<$ 0.35, 0 $<$ z $<$ 0.1, and 0.1 $<$ z $<$ 0.2 as shown in Fig. \ref{msfrall}. Respective comparison in stellar mass and redshift for each sub-sample can be seen in the inset histograms of Fig. \ref{msfrall}. The fitted coefficients are shown in Table \ref{msfrcoeff} and the differences in zero point between pair and control samples are shown in Table \ref{sfrzdiff}.

\begin{figure*}
\includegraphics[trim={0cm 7cm 0cm 0cm},clip,scale=0.48]{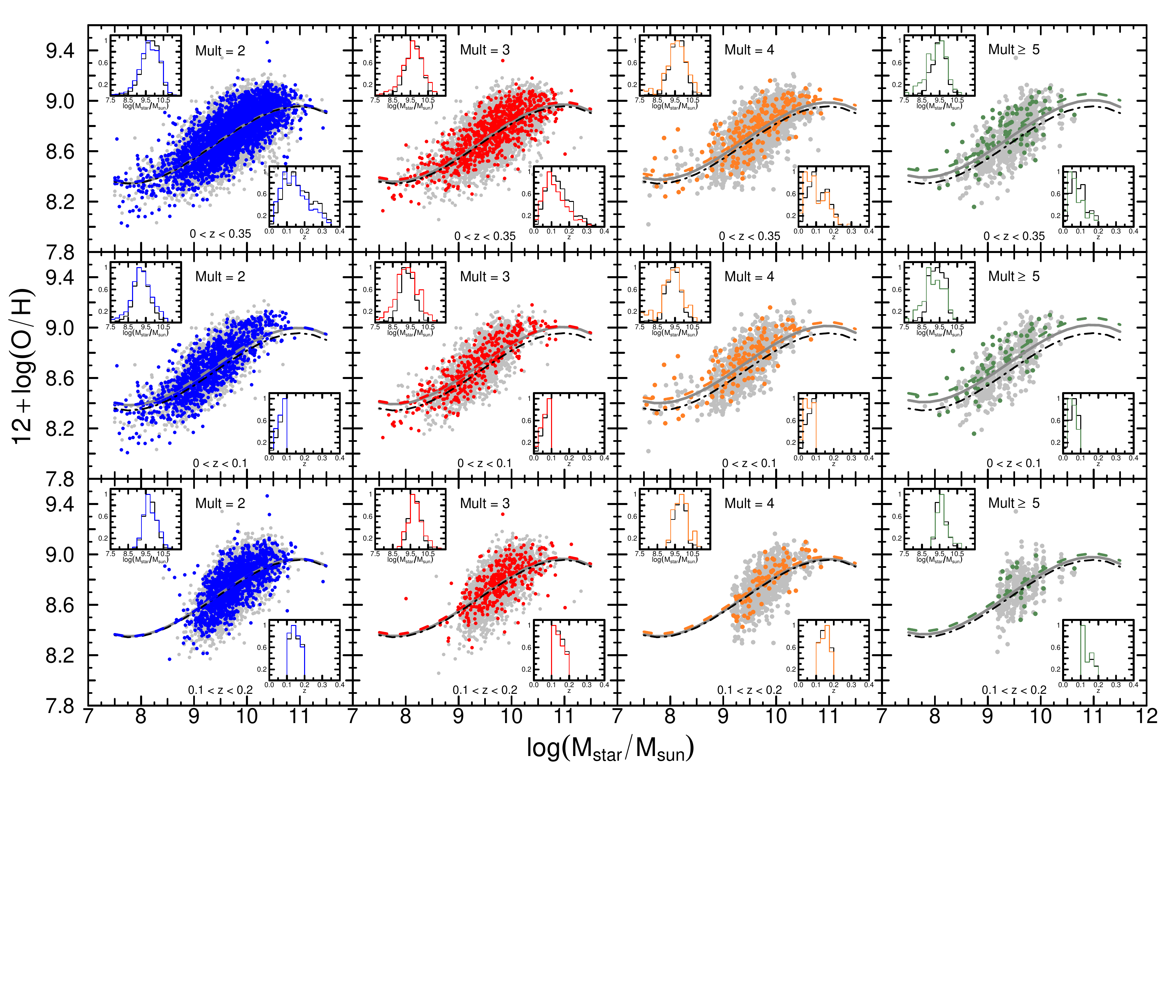}
\caption{The mass-metallicity relation (M-Z) for all galaxies. The colours and symbols are as for the previous figure.}
\label{mzall}
\end{figure*}

Similarly, for gas metallicities we use the M-Z relation and fit a third order polynomial. The difference between the zero points of both fits is our metallicity offset defined as $ \rm \Delta Z=log(Z)_{pair}-log(Z)_{control}$. The obtained fits and relationships are shown in Fig. \ref{mzall}. The fitted coefficients are shown in Table \ref{mzcoeff}, and its respective differences in zero point in Table \ref{sfrzdiff}.

\begin{figure*}
\includegraphics[trim={0cm 0.5cm 2cm 0cm},clip,scale=0.50]{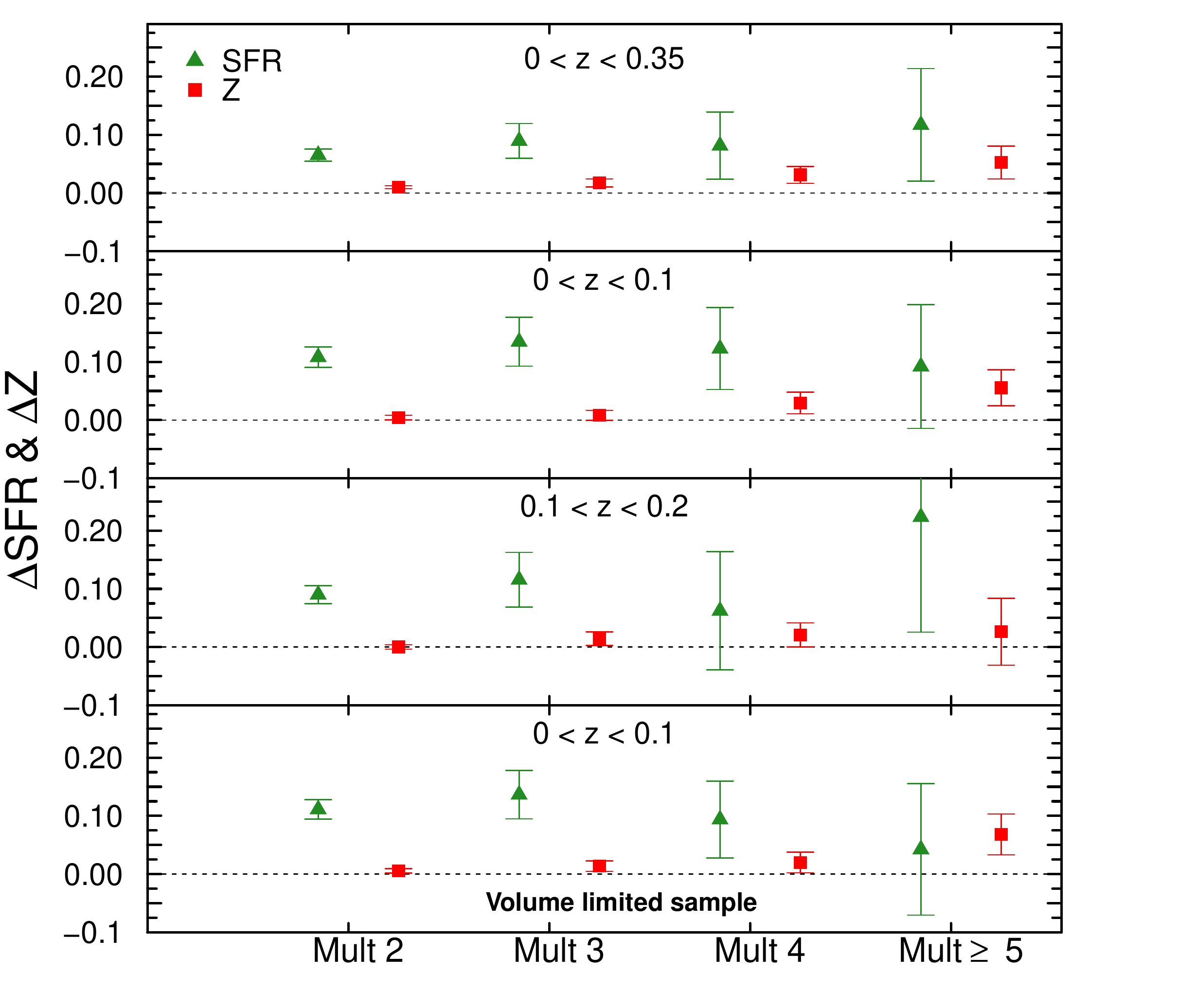}
\caption{Differences for SFR (triangles) and Z (squares) between galaxy pairs and control samples. The differences are shown for each sub-sample in multiplicity and redshift. Note that the fourth panel corresponds to the volume limited sample.}
\label{diffall}
\end{figure*}

A summary of our findings is given in Fig.\ref{diffall} and table \ref{sfrzdiff}. From top to bottom, the panels of Fig.\ref{diffall} show $\rm \Delta SFR$ for different redshift ranges, with general enhancements of $\rm \Delta SFR$ in all multiplicity cases. It can also be appreciated however, that $\rm \Delta SFR$ for M=4 and M$\geq$5 shown a high error, mainly due to low number statistics. 

In order to test the significance of our offsets, we apply an F-test, defined as the ratio between the variances of two populations, to assess whether they are statistically different. The results are summarized in Table \ref{tests}.
For all the redshift ranges, our $\rm \Delta SFR$ for M=2 has an enhancement of 0.06 dex (p-value $2.8\times 10^{-11}$), and we find a maximum enhancement of 0.12 dex (p-value $1.6\times 10^{-3}$) for M$\geq$5.

In general, galaxy pairs with M=2 and M=3 and for redshift 0 $<$ z $<$ 0.1 shows the most reliable results because for this, the associated deviations are relatively small.

Regarding metallicity, we find a slight $\rm \Delta Z$  enhancement of 0.01 dex for M=2 for all the redshift ranges, with a p-value of $6.4\times 10^{-17}$. Our maximum enhancement is 0.05 for M$\geq$5, although the p-value indicates this difference is not statistically significant.

Our analysis suggests that galaxy pairs produce an overall shift towards higher SFRs in the M-SFR relation. However, the gas metalliciy does not show substantial variations in the M-Z relation. Only higher multiplicities show a small increment in metallicity, although with high p-values, implying that a group environment may play a major role in producing these variations. 

To address the effect of higher multiplicities quantitatively, we consider the sample of single pairs (M=2) at 0$<$ z $<$ 0.35, and the differences found in Fig. 7. The inclusion of galaxies at higher multiplicity at the same redshift, and their respective control samples, would shift the zero point in the M-SFR relation by 27$\%$($\pm$4$\%$). Similarly, the effect in the M-Z relation would be of 4$\%$($\pm$1$\%$). Hence, the effect of higher multiplicities is important, specially for the SFR. 


To further address the effect of mass incompleteness as redshift increases, we repeat the same methodology with a volume limited sample of 4,393 SF galaxies in the redshift range 0 $<$ z $<$ 0.1, with a resulting mass range of 8.2 $<$ log( M$_{\star}$/ M$_\odot$) $<$ 10.9 (see Fig. \ref{GAMAz01box}). From such sample, we found 1,254 galaxy pairs and created a control sample of 4,253 galaxies (see also Table \ref{samples}). The results are shown in the fourth panel of Fig. \ref{diffall}. As can be seen, the most evident results are for M=4 and M$\geq$5, specifically there are $ \rm \Delta SFR $ differences of 0.03 and 0.05 dex respectively, with respect to the previous results in the same redshift range (second panel of Fig. \ref{diffall}). This suggests that the absence of the less massive galaxies in the volume limited sample drive the small differences in SFR at these multiplicities. In general, our methodology suggests that the use of control samples are an effective method to find differences in galaxy properties, and any bias generated by mass incompleteness should be small when control samples are used.

\begin{table*}
\centering
\begin{tabular}{cccccccc}
\toprule
\toprule
 &  & \multicolumn{6}{c}{F-tests (p-values)} \\
\cmidrule{3-8}
Multiplicity &  & \multicolumn{2}{c}{0 < z < 0.35} & \multicolumn{2}{c}{0 < z < 0.1} & \multicolumn{2}{c}{0.1 < z < 0.2} \\
\cmidrule{3-8}
 &  & SFR & Z & SFR & Z & SFR & Z \\
\midrule
\multirow{1}{*}{M=2} & & $2.8 \times 10^{-11}$ & $6.4 \times 10^{-17}$ & $6.3 \times 10^{-13}$ & $1.2 \times 10^{-9}$ & $2.1 \times 10^{-5}$ & $3.8 \times 10^{-3}$ \\

 \midrule
\multirow{1}{*}{M=3} & & $2.1\times 10^{-5}$ & $7.9\times 10^{-8}$ & $1.3 \times 10^{-5}$ & $4.1\times 10^{-6}$ & $4.9\times 10^{-5}$ & 0.22 \\
 
\midrule
\multirow{1}{*}{M=4} & & $6.2\times 10^{-4}$ & $4.6\times 10^{-3}$ & 0.01 & 0.08 & 0.02 & 0.3 \\
 
 \midrule
\multirow{1}{*}{M$\geq$5} & & $1.6\times 10^{-3}$ & 0.1 & 0.2 & 0.2 & 0.02 & 0.7 \\

\bottomrule
\end{tabular}
\caption{F-test for the differences of Fig. \ref{diffall}. We compare each galaxy pair samples vs their respective control sample for both, SFR and Z.}
\label{tests}
\end{table*}

\section{The effect of pair separation on SFR and Metallicity} \label{pairsec}

In this section we analyze the effect of the distance between paired galaxies in enhancing or decreasing the observed SFR and metallicity of galaxies. We define major pairs as pairs with a stellar mass ratio 0.5 $<$ $M_1$/$M_2$ $<$ 2; while pairs with a more discrepant stellar mass ratio are classified as minor pairs, in a similar way to \citet{Ellison08}.

For M$\geq$3, there are 2 or more galaxies associated to a single one, implying pairs of galaxies embedded in groups. This makes an analysis of pair separation difficult since there are 2 or more associated distances for each pair. It is also difficult to define a classification of major/minor pair since one galaxy could form a major pair with another, and at the same time, a minor pair with a second one. To constrain this problem, for pairs whose multiplicities are M$\geq$3, we take into account only the pair with the closest distance. 
Since we are selecting only the closest pairs for M$\geq$3, the number of galaxies is reduced. Hence, to increase our statistic we grouped all galaxy pairs with M$\geq$3. 

Our final sample in this section have galaxy pairs with relative velocities of $\Delta V$ $<$ 1000 km s$^{-1}$ and separations up to 100 ${\rm h}^{-1}$ kpc, and consists of 1,374 major pairs and 3,265 minor pairs.

The methodology used to analyze the role of pair separation in the SFR and Z is described below, and it is applied to both, major and minor pairs. First, galaxy pairs are grouped in bins of 10 ${\rm h}^{-1}$ kpc from 0 to 100 ${\rm h}^{-1}$ kpc. Then, control samples are generated for paired galaxies in each bin following the methodology described in \S 3.

Next, we compute the median SFR and Z of each bin for both, pairs and control galaxies. We apply the definition of offset to SFR and Z as in the previous section.
This process is repeated for major and minor pairs for the redshift range 0 $<$ z $<$ 0.35. Additionally, the results for the redshift range 0 $<$ z $<$ 0.1 are shown in Appendix \ref{pairsepappendix}. In each sub-sample, we show separately the most and least massive galaxy of the pair.

Finally, error bars are computed using the standard error (SE) of the median \citep{abu2009confidence} in the galaxy pair sub-samples according to:

\begin{equation}
\rm \sigma_{M\;=\;1.253\;}\left[\frac{\sqrt{\frac{\Sigma{(X_i-\mu)}^2}{n-1}}}{\sqrt n}\right]
\end{equation}

\begin{figure*}
\center
\includegraphics[trim={0 6cm 0 0cm},clip,scale=0.5]{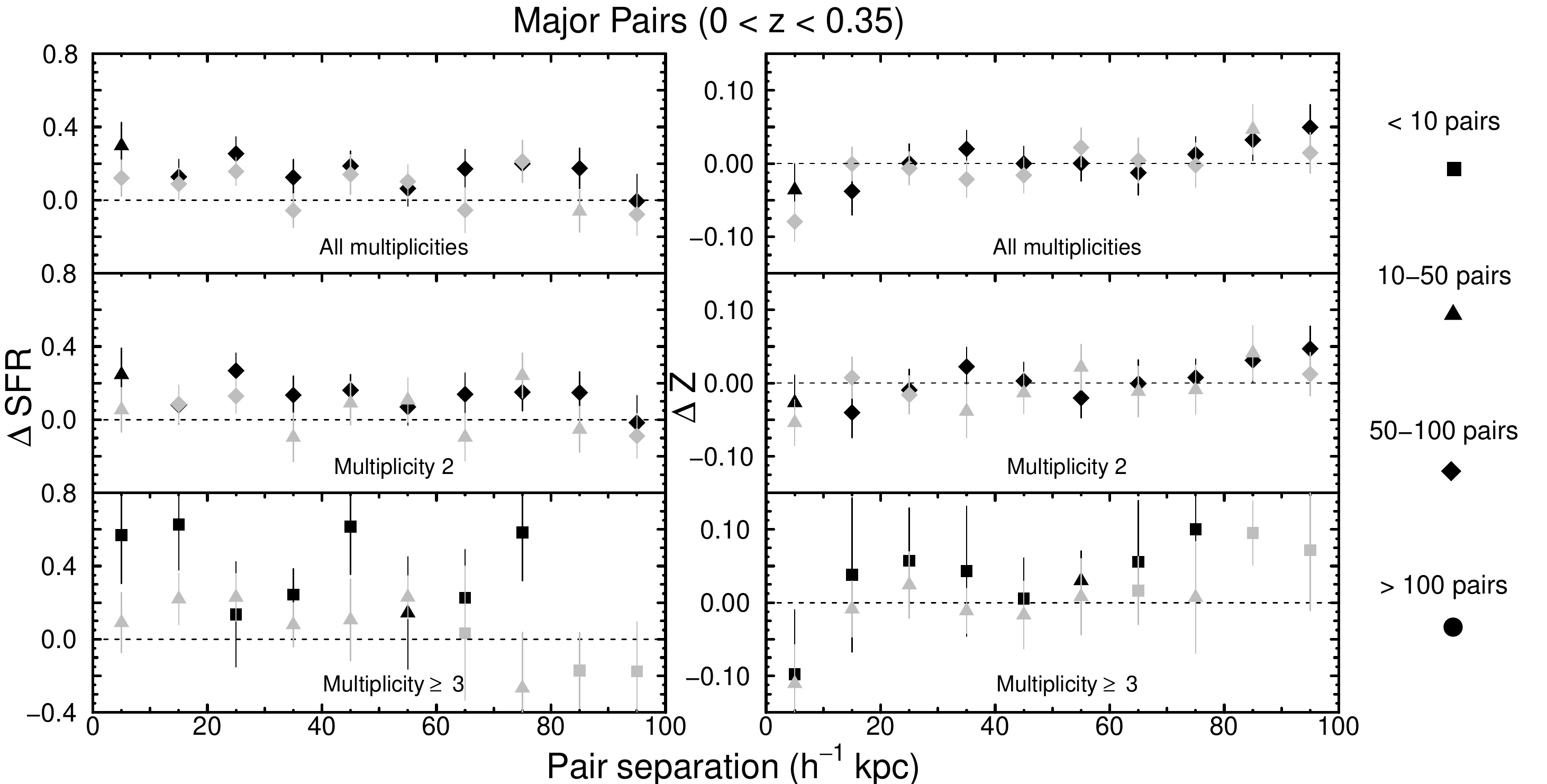}
\caption{Differences in SFR (left panel) and Z (right panel) as a function of pair separation for major pairs. Black symbols correspond to the most massive member of the pair, while gray symbols to the least massive. Symbols represent the number of galaxies in each bin as indicated on the right hand side of the graph. We require at least 5 galaxies to plot a symbol in each bin so that empty bins does not satisfy the condition.}
\label{allmajo}
\end{figure*}
\begin{figure*}
\center
\includegraphics[trim={0 6cm 0 0cm},clip,scale=0.5]{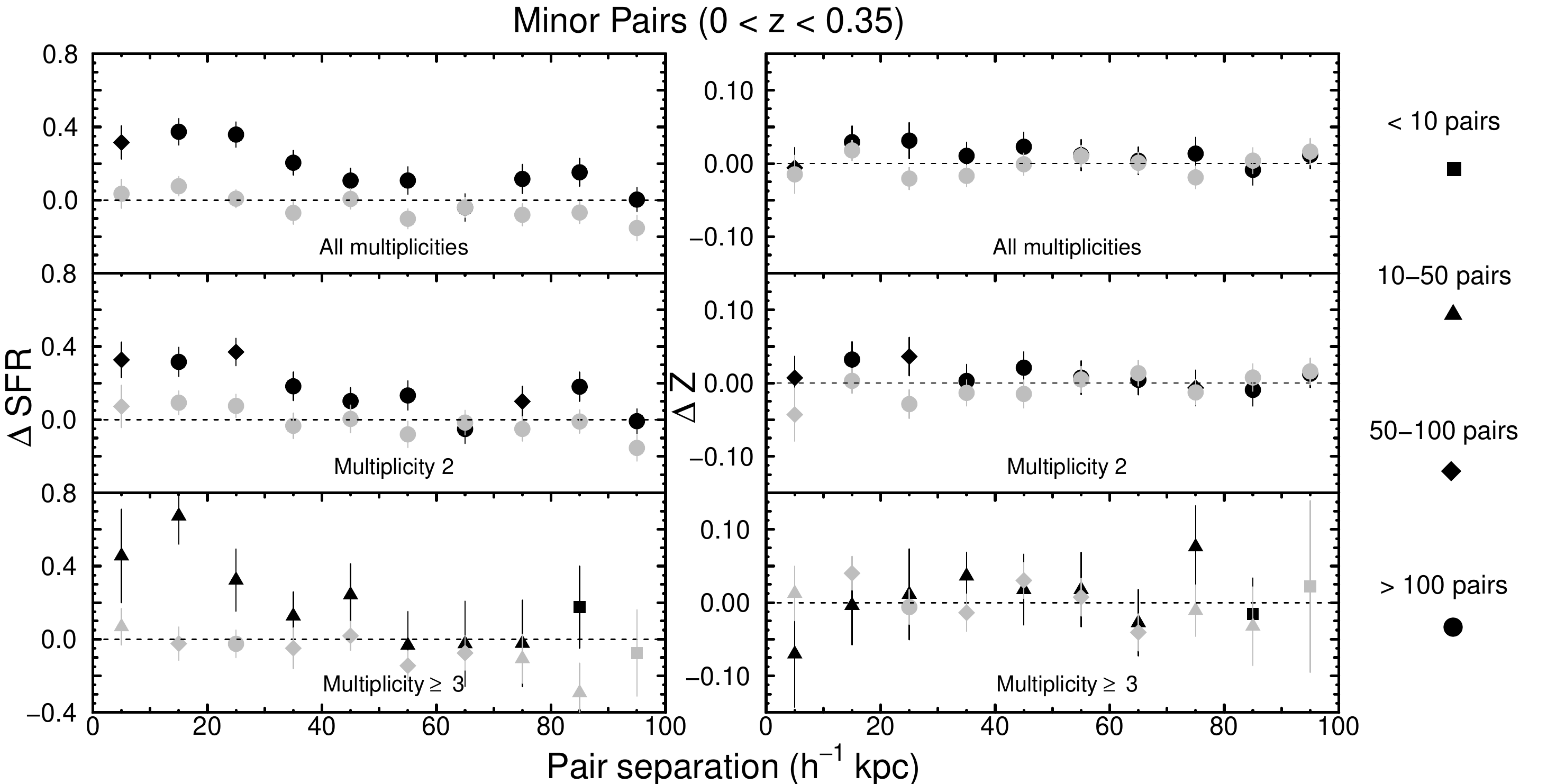}
\caption{Same symbols as in in Fig. \ref{allmajo} for minor pairs.}
\label{allmino}
\end{figure*}

In general, we observe enhancements  as a function of pair separation in SFR and a small decrement in Z for the closest pairs. The more evident tendencies are found when data at all multiplicities are considered (top panels in Figures \ref{allmajo} and \ref{allmino}). The most discrepant results, which also show the largest error bars, are found for M$\geq$3.

A Wilcoxon-Mann-Whitney test (U-test) is applied in this section to confirm the significance of our results. This test is specifically designed to compare median values. For major pairs with M=2 (Fig. \ref{allmajo}), we find enhancements of up to 0.30 dex in $ \rm \Delta SFR $ for the most massive galaxy of the pair, with a p-value of $6\times 10^{-3}$. However, the enhancements in SFR oscillate with pair separation with no clear pattern, showing even a few negative values for the least massive member of the pair. When M$\geq$3, the data show a high scatter, although the data are consistent with an enhancement in SFR when the pair separation is lower than $\sim$50 ${\rm h_{70}}^{-1}$ kpc.

For $ \rm \Delta Z $ of all multiplicities, we find that the metallicity of the closest pairs in all the range of redshift show small decrements of $\sim$-0.08 dex (p-value of $2.2\times 10^{-4}$). As the distance between galaxies increases, $ \rm \Delta Z $ oscillates around zero and shows a slight increment for larger separations. This tendency is more clear when all multiplicities are considered, and with a larger scatter for M$\geq$3 (Fig. \ref{allmajo}).

On one hand, minor pairs show a more distinctive pattern, with an enhancement in SFR for the most massive galaxy of the pair of $\sim$0.37 dex (all multiplicities, top panel of Fig. \ref{allmino}) and up to 0.67 dex (M$\geq$3, bottom panel of Fig.\ref{allmino}), both for pair separations up to $\sim$30 ${\rm h}^{-1}$ kpc with a p-value less than $2.2\times 10^{-4}$. On the other hand, the metallicity shows only minor variations, with small  $ \rm \Delta Z $ enhancements of 0.02 dex (all multiplicities) and 0.04 (M$\geq$3) for the least massive galaxy of the closest pairs ($<$ 30 ${\rm h}^{-1}$ kpc), although the U-test do not indicate these values are significant.

In general, we find that both galaxies in major pairs show enhancements in SFR. The strongest enhancements in SFRs are found for the most massive galaxy in minor pairs at separations $<$ 30 ${\rm h}^{-1}$ kpc. For gas metallicities, we find the strongest decrements in major pairs at close distances ($< $10 ${\rm h}^{-1}$ kpc)


\section{The effect of dynamics on SFR and Metallicity}\label{dyneff}

In this section we provide a new approach to measure the variation of SFR and metallicity based on dynamics. The collapsing of any overdensity can be described by self-similar secondary infall and accretion, as proposed by \citet{1985ApJS...58...39B} for an Einstein-de Sitter universe. This scenario follows the growth of overdensities as they detach from the cosmic flows at a maximum radius called the first turn-around radius ($r_{1t}$), as they collapse for their first time. After crossing, overdensities will bounce expanding to maximum radius, smaller than $r_{1t}$, the second turn-around radius ($r_{2t}$), turning around and collapsing for a second time. At $r_{1t}$ and $r_{2t}$ overdensities are neither expanding nor collapsing, generating discontinuities in density and velocity dispersion. \citet{2015AJ....149...54T} suggested that both radii can be recovered from observations. We can approach the dynamics of galaxy pairs at least in a statiscal manner considering $v_{sep}$ and the formalism developed by \citet{2015AJ....149...54T}. 

\citet{2015AJ....149...54T} proposed a more general definition of groups by considering associations, groups, and clusters of galaxies. This approach can be applied to any overdensity included pairs, or associations, such as the Milky Way and Andromeda. It was found that the line-of-sight velocity dispersion defined by $\sigma_p=\sqrt{\Sigma_{i}(v_{i}-\overline{v})/N}$ (the population standard deviation of $N$ velocities $v_i$, where $\overline{v}$ is the average velocity) and the projected  second turn-around radius $R_{2t}$ are tightly correlated, admitting the following parametrization: 
\begin{equation}
    \frac{\sigma_{p}}{R_{2t}}=(343\pm 7)\,h_{} \,\mathrm{km\,s^{-1}\,Mpc^{-1}}\label{eqn-1}.
\end{equation}

For a pair of galaxies $\sigma_p=\left(\lvert{v_{sep}}\rvert / \sqrt{2} \right)$. Equation \ref{eqn-1} was established for $70\le\; {\sigma_{p}}/\mathrm{km\,s^{-1}} \le 1000$ \citep{2015AJ....149...54T}. It can be shown that the virial mass can be calculated by the following formula \citep{2015AJ....149...54T}: 
\begin{equation}
    \mathcal{M}_v=2.1\times 10^{6} \, \sigma_{p}^{3}\,h^{-1}_{}\;M_{\sun}.
\end{equation}

An estimation of the mass of the systems gives us a proxy for the size of the halo where the pair is embedded, which gives us a complementary indicator of the environment besides multiplicity.

In the self-similar secondary infall scenario, overdensities with the larger masses would collapse faster \citep{1985ApJS...58...39B}; hence, pairs in most massive overdensities may have experienced previous episodes of enhanced star formation induced by tidal encounters during their crossings \citep{2019ApJ...886L...2L}. An enhancement in the galaxy metallicity should result afterwards. The lowest velocity dispersion observed for our sample of GAMA pairs is $\sigma_{p}\sim50\; \mathrm{km\,s^{-1}}$, which corresponds to $R_{2t}\sim145 \; \mathrm{kpc}$. This means that all the pairs selected in this study are inside their second turn-around radius. For $\sigma_{p} > 300$ km s$^{-1}$ the pair might be embedded in a 
$\mathcal{M}> 6\times10^{13}\;h^{-1}_{}\;M_{\sun}$ halo; hence, pairs with high multiplicity and high $\sigma_p$ are likely to be found in groups or poor clusters of galaxies. 

\begin{figure*}
\center
\includegraphics[trim={0 6cm 0 0cm},clip,scale=0.5]{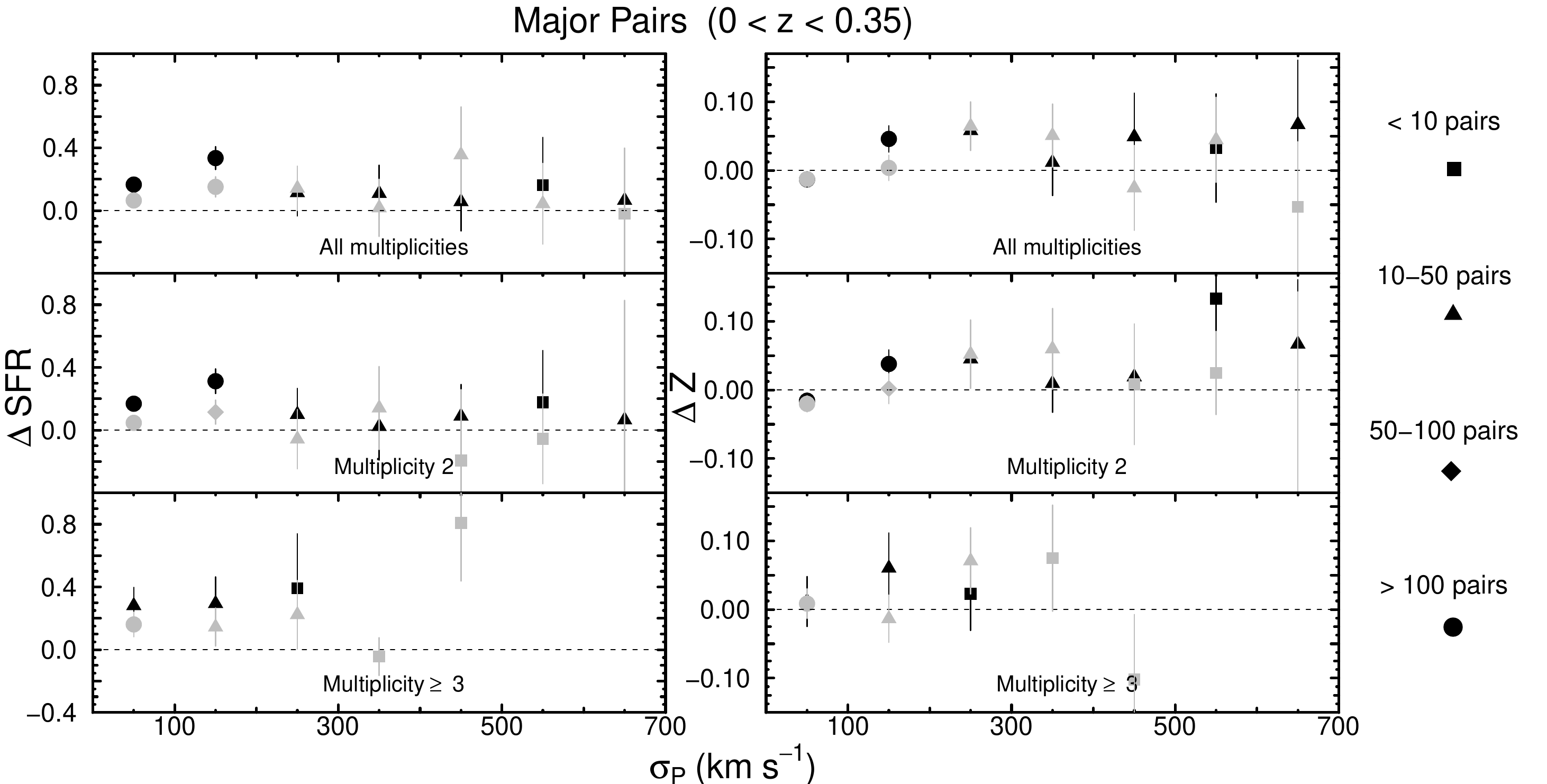}
\caption{Differences in SFR (left panel) and Z (right panel) as a function of velocity dispersion for major pairs. The same notation as in  Fig \ref{allmajo} is employed: black symbols correspond to the most massive member of the pair, while gray symbols to the least massive. Symbols represent the number of galaxies in each bin as indicated on the right hand side of the graph. In each diagram the multiplicity is indicated.  We require at least 5 galaxies to plot a symbol in each bin so that empty bins does not satisfy the condition.}
\label{velo1}
\end{figure*}

\begin{figure*}
\center
\includegraphics[trim={0 6cm 0 0cm},clip,scale=0.5]{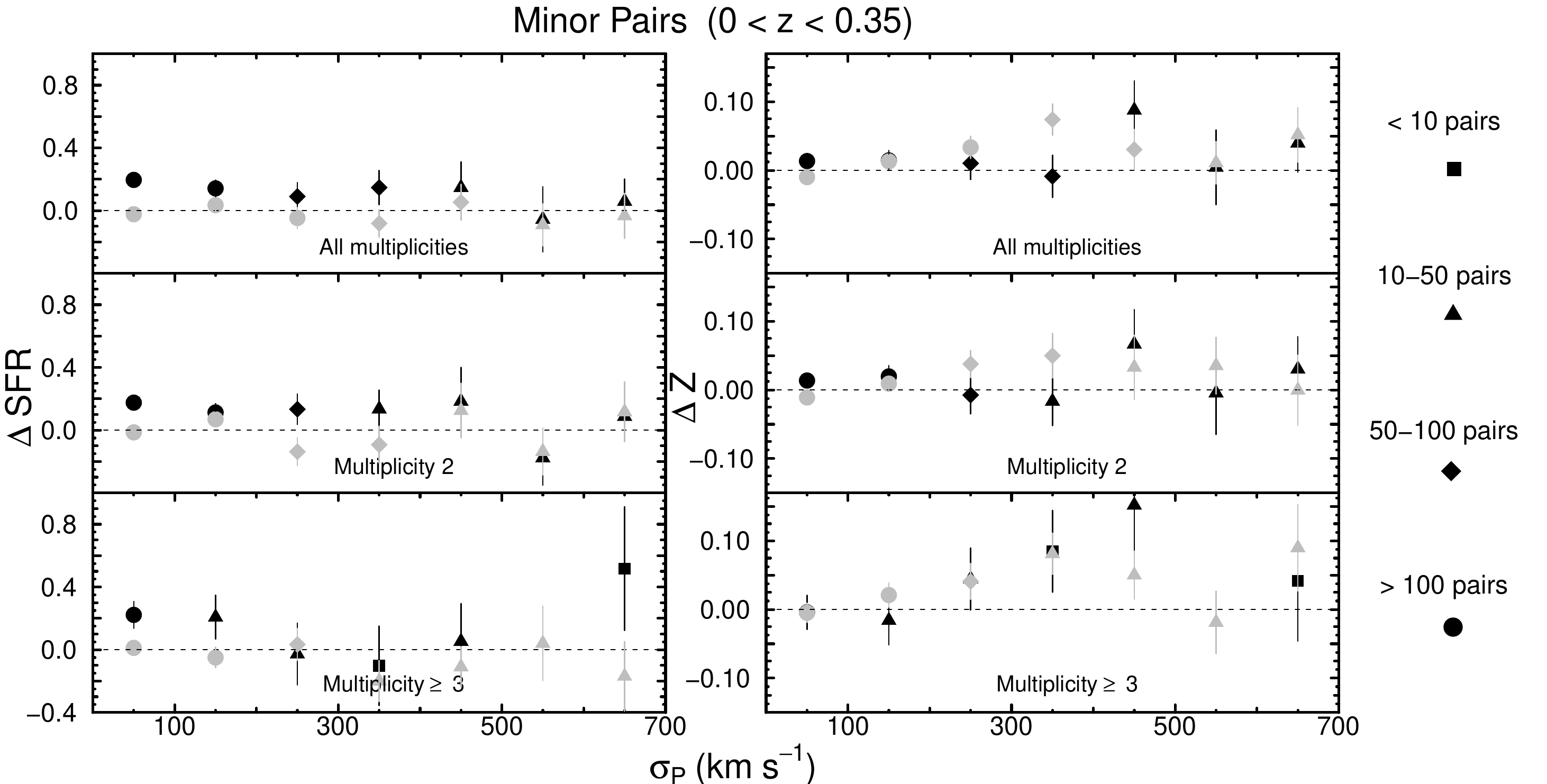}
\caption{Differences in SFR (left panel) and Z (right panel) as a function of velocity dispersion for minor pairs. The symbols are the same as in Fig. \ref{velo1}.}
\label{velo2}
\end{figure*}

We explore the effects of the velocity dispersion on the SFR and Z for pairs according to multiplicity. For the generation of Figures \ref{velo1} and \ref{velo2}, we use our sample in the redshift range 0 $<$ z $<$ 0.35. A U-test is also applied in this section to asses the significance of our results. For the more massive galaxy of major pairs of all multiplicities, we find a $ \rm \Delta SFR $ enhancement up to 0.33 dex with a p-value of $7\times 10^{-6}$ for a $\sigma_{p} < 200\; \mathrm{km\,s^{-1}}$; for the less massive galaxy an enhancement of 0.15 dex (p-value 0.03) is found for the same $\sigma_{p}$ range. Regarding metallicity, we find $ \rm \Delta Z $ enhancements up to 0.07 dex for the most and least massive galaxy of major pairs, with a p-value less than 0.05 for a $\sigma_{p} > 100\; \mathrm{km\,s^{-1}}$, while for $\sigma_{p} < 100\; \mathrm{km\,s^{-1}}$ $, \rm \Delta Z $ do not show variations.

In general, we find  stronger enhancements for $ \rm \Delta Z $ when the velocity dispersion is considered, in contrast with the pair separation analysis of Sect. \S \ref{pairsec}.

\section{Discussion}

Many properties of galaxies such as the SFR and Z, are affected by their environment. Due to its complex nature and potentially strong effect on the evolution of galaxies, many analyses have targeted this problem from the observational and theoretical point of view.

The sample analyzed in this paper consists of 76.3\% of galaxies with multiplicity M=2, or simple pairs, and 23.7\% of galaxies with multiplicities M$\geq$3. These proportions are similar for the different redshift ranges studied. Previous studies have analyzed properties of galaxy pairs as a function of pair separation without considering galaxies at high multiplicities \citep[e.g,][]{Ellison08,2010MNRAS.407.1514E,2013MNRAS.435.3627E}.

In \S \ref{multsec} we analyze the M-SFR and M-Z relations of galaxy pairs with different multiplicities. When all redshifts are considered (top panel of Fig. \ref{diffall}), our data suggest SFR enhancements with a median value of 0.08 dex. When we take bins of redshift, our data suggest similar enhancements between simple pairs and higher multiplicities, with SFR enhancements of up to 0.23 dex for galaxies at M$\geq$5 (although with very high uncertainties). The enhancements in SFR we find are consistent with the paradigm of an enhanced SFR for galaxy interactions of isolated compact groups (CGs) as mention in \citet{2012MNRAS.423.2690S}. Scudder et al. point out the importance of the environment and the differences of SFR in isolated and embedded CGs. They also show how isolated CGs could have values of enhanced SFR ($\sim$0.08-0.24 dex) similar to previous studies of galaxy-galaxy interactions \citep[e.g,][]{Ellison08,2010MNRAS.407.1514E,2015MNRAS.452..616D}.

Regarding gas metallicity, we find only a slight increase of $ \rm \Delta Z $ $\sim$0.05 for multiplicity M$\geq$5. This result is comparable with \citet{2009MNRAS.396.1257E}, who find Z enhancements of 0.04-0.06 dex in high density environments. As pointed out in \citet{2009MNRAS.396.1257E} and \citet{2012MNRAS.423.2690S},  Z enhancements could be reached by galaxies in different environmental scenarios, since it is likely that pairs at high multiplicities are embedded in, or form groups of galaxies. 

The environments in which interactions are happening is an important parameter to consider, since it could directly have repercussions in SFR, Z or other properties \citep[e. g.,][]{2007MNRAS.381..494O,2009MNRAS.399.1157P,2010MNRAS.407.1514E,2012A&A...539A..46A}. In certain environments, some known effects could prevail more than others, such as harassment, ram preasure stripping, starvation or strangulation \citep{2012MNRAS.423.2690S}

In \S \ref{pairsec} we analyze the effects of minor and major pairs as a function of pair separation, our more reliable results are for pairs with M=2. Some authors \citep[e.g., ][]{Ellison08, 2012MNRAS.426..549S,2015MNRAS.452..616D} report enhancements in SFR of up to 0.24 dex for galaxy pairs at close separations ~< 40 ${\rm h_{70}}^{-1}$ kpc, especially for pairs of similar stellar mass. Our results of SFR enhancements show median values of 0.17 dex and 0.13 dex for the most massive galaxy of major and minor pairs, respectively (first panel of Figs. \ref{allmajo} and \ref{allmino}) which agree with the aforementioned studies. This is consistent with a scenario where the interaction of close galaxies induces gas inflows that enhances the SFR; the efficiency of such SFR is determined by the properties of each galaxy in the interaction. Presumably, there is more efficiency in the SFR for interacting galaxies when they are in major pairs \citep{2007A&A...468...61D,Ellison08,2007MNRAS.381..494O,2012MNRAS.426..549S,2011BAAA...54..309M,2012A&A...539A..45L}, although it is also known that minor interactions have considerable importance \citep{2006AJ....132..197W,2014MNRAS.440.2944K}. Moreover, \citet{2015MNRAS.452..616D} postulate that for galaxy pairs in different stages of interaction, the least massive galaxy could be massive enough to retain its gas and continue its star formation (major pairs), or be stripped out of its gas, quenching the star formation (minor pairs). From our results, the more extreme differences are observed as SFR enhancement for the most massive galaxy in minor pairs, implying a possible gas inflow of gas, in agreement with  \citet{2015MNRAS.452..616D}.

It was reported in \citet{2013MNRAS.435.3627E} that the SFR is enhanced by a factor of a few in the pre-merger phase of interacting galaxies; however, in the post-merger phase, there could be enhancements of a factor $\sim$3.5. Our results in $ \rm \Delta SFR $ goes from 0.3 to 0.6 dex, which is in agreement with the two scenarios previously mentioned. 

As for metallicity, we find small decrements of up to $\sim$0.1 dex for close galaxy pairs, in agreement with different studies \citep[e.g,][]{Ellison08,2012MNRAS.426..549S,2013MNRAS.435.3627E}. The decrements in metallicity are precisely associated to gas inflows of pristine gas that dilutes the metals that are already there, this effect is evidenced by enhancements in SFR, flattening in the gradients of gas metallicity \citep{2010ApJ...721L..48K, 2019MNRAS.482L..55T}, and by abnormal decrements in gas metallicity in SF galaxies \citep{2019ApJ...872..144H}.

Also, \citet{2013MNRAS.435.3627E} point out that post-merger systems reach a 0.1 dex decrement in metallicity, which is consistent with the 0.12 dex decrement we find for our major pairs at 0 $<$ z $<$ 0.1. This means that the stage of the interaction in galaxies could change the associated decrements.

We find some enhancements in SFR of up to 0.1 dex for galaxies with separations larger than 30 ${\rm h_{70}}^{-1}$ kpc, and decrements in metallicity up to $\sim$0.05 dex, similar to \citet{2012MNRAS.426..549S}, who find a SFR enhancement of 0.12 dex out to 80 ${\rm h_{70}}^{-1}$ kpc, and metallicity decrements from 0.01-0.05 dex.

In \S \ref{dyneff} we analyze the SFR and Z as a function of $\sigma_p$. The $\rm \Delta SFR$ shows small enhancements for major pairs when $\sigma_p < \mathrm{200 km s^{-1}}$. This is expected, if star formation is induced by tidal interaction, tidal heating is more efficient in slow encounters \citep[e.g.,][]{2008LNP...740...71A}. Therefore, enhanced SFR is expected to be found in close pairs with $< \,40 {\rm h_{}}^{-1}$ kpc and low $\sigma_p < 200$ km s$^{-1}$ both in major and minor pairs at all multiplicities. 

Minor and major pairs do show enhanced metallicities for most multiplicities, although pairs with $\sigma_p < 100$ show minor variations with respect to the control sample. The most dramatic increments in $\rm \Delta Z$ are observed for minor pairs with high multiplicity and $\sigma_p > 200$ (Figure \ref{velo2}). Compact groups have a chance to cross at least once during a Hubble time \citep{2019ApJ...886L...2L}.
During the first crossing galaxy encounters may induce starbursts which could use a large fraction of the galaxy's gas. As a result, the increment in metallicity that is observed in minor pairs of high multiplicity, reveals that star formation is happening in galaxies whose interstellar media have been pre-enriched. We therefore suggest that minor pairs of  high multiplicity reside in compact groups or loose  groups; however, since we have followed the metallicity and SFR using emission lines, the lack of major pairs in high multiplicity environments suggests that these pairs are found in richer environments such as clusters of galaxies.

Our results are also in agreement with simulations. Different authors have found SFR enhancements for galaxy interactions, even at different stages. From \citet{2007A&A...468...61D}, it is known that major interactions and merger systems could produce strong starbursts in the central regions of galaxies (although this does not always occur). They also report how galaxy interactions can increase the SFR for galaxy pairs in different levels, finding SFR enhancements from 1.5-2 times to 20 times, this last value typical of starburst galaxies.   

Moreover, studies have also focused on explaining how SFR enhancements at wide separations could be reached. For instance,  by simulating a galaxy pair interaction, \citet{2012MNRAS.426..549S} show how after the first encounter in a pair separation of 60 kpc, a SFR enhacement of a factor of 3.9 could be reached. \citet{2013MNRAS.433L..59P} also probe that high SFR enhancements can be produced in simulations with realistic orbit parameters, at a distance of 60 kpc, they find a SFR enhancement factor of 1.8. For the case of metallicities, \citet{2012MNRAS.426..549S} find a  decrement of 0.10 dex after the first encounter in a pair separation of 60 kpc.

It is important to highlight that new integral field unit (IFU) studies of interacting systems, have found important evidence of variations in SFR and Z in non central regions of the galaxies \citep{2019ApJ...872..144H,2019ApJ...881..119P,2019MNRAS.482L..55T,2019ApJ...886L...2L}. Nevertheless, these authors agree that it is in the center of galaxies where larger differences would be appreciated. In practice, this could change the common values estimated by large surveys that use fiber spectroscopy, such as the SDSS and GAMA surveys.

\section{Conclusions}
We present an analysis of star formation rate and metallicity for a sample of 4,636 SF galaxy pairs using the GAMA survey. Our pairs were taken from the GAMA Galaxy Group catalogue (G$^3$C) and all our spectroscopic information from the SpecLineSFRv05 GAMA catalogue. 

Our analysis focuses on three main topics, the variations of SFR and Z of galaxy pairs as a function of multiplicity, pair separation and velocity dispersion. Our main sample of pairs was divided in sub-samples of multiplicities and redshift bins. For each sub-sample, we created control samples from field galaxies by finding matches in stellar mass and redshift.

In the first part of our analysis, we focused on the effect of galaxy pairs in the M-Z and M-SFR relation. From this analysis, we arrive at the following conclusions:
\begin{enumerate}
   \item The SFR shows an enhancement in the M-SFR relations for galaxies in pairs in all our sub-samples of multiplicity and redshift.
   \item Our data considering all redshift ranges suggest that the SFR enhancement increases with multiplicity, which is likely an effect of pairs embedded in groups. Also, higher multiplicities could produce a bias towards higher SFRs of up to $\sim$27$\%$ in studies of pairs that do not take into account multiplicity.
   \item Gas metallicity remains almost invariant in the M-Z relation for galaxy pairs. However, galaxies in pairs with M$\geq$5 do show a small increment in metallicity, which is likely caused by the pairs being part of group of galaxies. Thus the effect of pairs at high multiplicity would have a weak effect in the M-Z relation of $\sim$4$\%$.
\end{enumerate}

For the second part of our analysis, we analyze changes in SFR and Z as a function of pair separation. Our main conclusions of this part of the analysis are given below:

 \begin{enumerate}
   \item Our sample indicates SFR enhancements for both, minor and major pairs, with respect to their control samples. The strongest enhancements are observed for the most massive galaxy in minor pairs, and for distances lower than 30 ${\rm h_{70}}^{-1}$ kpc.
   \item The observed enhancements in SFR are stronger when M$\geq$3, in both minor and major pairs, at close separations.
   \item Gas metallicities show a significant decrement for the closest pairs in both minor and major pairs. For larger distances the differences are negligible.
   \item For major pairs, both galaxies in the pair show very similar enhancements in SFRs of up to $\sim$0.30 dex. Gas metallicities show a decrement, for the closest pairs, of $\sim$0.08 dex for the less massive galaxy in the pair.
   \item For minor pairs, the most massive galaxy shows SFR enhancements of up to 3 times with respect to the control sample, while the least massive galaxy shows minor variation. In this case, the metallicity does not exceed 0.05 dex decrement for the closest pairs, although with a high p-value, and show negligible variations at any other distance.
\end{enumerate}

For the third part of our analysis, we adopt an analysis on dynamics, specifically in the velocity dispersion of the pairs. Our main conclusion are given below:

\begin{enumerate}
   \item Small enhancements in SFR are observed in major and minor pairs with low velocity dispersion $(\sigma_p < 250\, \mathrm{km\, s^{-1}})$. 
   \item Enhanced metallicity was found in minor pairs with high velocity dispersion $(\sigma_p < 250\, \mathrm{km\, s^{-1}})$ and high multiplicity. We suggest that these pairs should be found in compact groups or in loose groups.
   \item Contrary to pair separation, the velocity distance interpreted as velocity dispersion shows a dependence with metallicity for major and minor pairs. 
\end{enumerate}



\section*{Acknowledgments}

GAMA is a joint European-Australasian project based around a
spectroscopic campaign using the Anglo-Australian Telescope. The
GAMA input  is based on data taken from the Sloan Digital Sky Survey and the UKIRT Infrared Deep Sky Survey. Complementary imaging of the GAMA regions is being obtained by
a number of independent survey programs including GALEX MIS,
VST KiDS, VISTA VIKING, WISE, Herschel-ATLAS, GMRT and
ASKAP providing UV to radio coverage. GAMA is funded by the
STFC (UK), the ARC (Australia), the AAO, and the participating
institutions. The GAMA website is http://www.gama-survey.org/.
MALL is a DARK-Carlsberg Foundation Fellow (Semper Ardens project CF15-0384).
We acknowledge support from CONACYT (studentship 454033 2017-2019), CONCYTEP (student grant 2018), INAOE-Astrophysics (studentship 2019) and the CONACYT SNI-III studentship given by Sabino Chávez-Cerda (studentship 17906).

\section*{Data Availability}

The data underlying this article are or soon will be made available through GAMA public data releases; see http://gama-survey.org/. The derived results shown in this paper will be shared on reasonable request to the corresponding author.



\bibliographystyle{mnras}
\bibliography{Bibliografia.bib} 




\appendix

\section{Differences and coefficients from M-SFR and M-Z fittings.}

The differences found in $ \rm \Delta SFR $ and $ \rm \Delta Z $ from Fig. \ref{diffall} (\S 3) are listed in the table \ref{sfrzdiff}.
 The coefficients of the polynomial fits from the same section are shown here. Table \ref{msfrcoeff} corresponds to  Fig. \ref{msfrall} and Table \ref{mzcoeff} to Fig. \ref{mzall}.

\begin{table*}
\centering
\begin{tabular}{@{}clcccccccc@{}}
\toprule
\toprule
 &  & \multicolumn{6}{c}{Differences} \\ 
\cmidrule{3-8}
Multiplicity &  & \multicolumn{2}{c}{0 \textless{} z \textless{} 0.35} & \multicolumn{2}{c}{0 \textless{} z \textless{} 0.1} & \multicolumn{2}{c}{0.1 \textless{} z \textless{} 0.2} \\
 \cmidrule{3-8}
 &  & SFR & Z & SFR & Z & SFR & Z \\ \midrule
M=2 &  & 0.07 $\pm$ 0.01 & 0.01 $\pm$ 0.01 & 0.11 $\pm$ 0.02 & 0.01 $\pm$ 0.01 & 0.09 $\pm$ 0.01 & 0.01 $\pm$ 0.01 \\
M=3 &  & 0.09 $\pm$ 0.02 & 0.02 $\pm$ 0.01 & 0.14 $\pm$ 0.03 & 0.01 $\pm$ 0.01 & 0.12 $\pm$ 0.05 & 0.01 $\pm$ 0.01 \\
M=4 &  & 0.08 $\pm$ 0.04 & 0.03 $\pm$ 0.01 & 0.12 $\pm$ 0.06 & 0.03 $\pm$ 0.01 & 0.06 $\pm$ 0.10 & 0.01 $\pm$ 0.01 \\
M$\geq$5 &  & 0.12 $\pm$ 0.08 & 0.05 $\pm$ 0.02 & 0.09 $\pm$ 0.09 & 0.06 $\pm$ 0.03 & 0.22 $\pm$ 0.19 & 0.03 $\pm$ 0.05 \\ \bottomrule
\end{tabular}
\caption{Values of the offsets $\rm \Delta SFR$ and $\rm \Delta Z$ in the different ranges of redshift.}
\label{sfrzdiff}
\end{table*}

\begin{table*}
\centering
\begin{tabular}{cccccc}
\toprule
\toprule
\multicolumn{6}{c}{Coefficients for the M-SFR relation} \\
\midrule
Redshift & Coefficients & M=2 & M=3 & M=4 & M$\geq$5 \\
\midrule
\multirow{2}{*}{0 < z < 0.35} & $c$ & -9.15 $\pm$ 0.01 & -9.20 $\pm$ 0.01 & -9.25 $\pm$ 0.01 & -9.26 $\pm$ 0.01 \\
 & $c_p$ & -9.08 $\pm$ 0.01 & -9.11 $\pm$ 0.02 & -9.17 $\pm$ 0.03 & -9.15 $\pm$ 0.05 \\ \midrule
\multirow{2}{*}{0 < z < 0.1} & $c$ & -9.28 $\pm$ 0.01 & -9.29 $\pm$ 0.01 & -9.32 $\pm$ 0.01 & -9.29 $\pm$ 0.02 \\
 & $c_p$ & -9.17 $\pm$ 0.01 & -9.16 $\pm$ 0.03 & -9.20 $\pm$ 0.04 & -9.20 $\pm$ 0.06 \\ \midrule
\multirow{2}{*}{0.1 < z < 0.2} & $c$ & -9.10 $\pm$ 0.01 & -9.20 $\pm$ 0.01 & -9.19 $\pm$ 0.01 & -9.22 $\pm$ 0.02 \\
 & $c_p$ & -9.19 $\pm$ 0.01 & -9.09 $\pm$ 0.02 & -9.13 $\pm$ 0.04 & -8.99 $\pm$ 0.12 \\
\bottomrule
\end{tabular}
\caption{Coefficients of the linear fit for the M-SFR relation. The coefficients are given for each sub-sample in multiplicity and redshift, $c$ is the zero point for the control sample, while $c_p$ is the zero point of the galaxy pairs. The slope is taken from the fiducial fit shown in Fig.\ref{GAMAfid}}
\label{msfrcoeff}
\end{table*}

\begin{table*}
\centering
\begin{tabular}{cccccc}
\toprule
\toprule
\multicolumn{6}{c}{Coefficients for the M-Z relation} \\
\midrule
Redshift & Coefficients & M=2 & M=3 & M=4 & M$\geq$5 \\
\midrule
\multirow{2}{*}{0 < z < 0.35} & $c$ & 37.15 $\pm$ 0.01 & 37.16 $\pm$ 0.01 & 37.18 $\pm$ 0.01 & 37.20 $\pm$ 0.01 \\
 & $c_p$ & 37.16 $\pm$ 0.01 & 37.18 $\pm$ 0.01 & 37.21 $\pm$ 0.01 & 37.25 $\pm$ 0.01 \\ \midrule
\multirow{2}{*}{0 < z < 0.1} & $c$ & 37.19 $\pm$ 0.01 & 37.20 $\pm$ 0.01 & 37.20 $\pm$ 0.01 & 37.27 $\pm$ 0.02 \\
 & $c_p$ & 37.19 $\pm$ 0.01 & 37.20 $\pm$ 0.01 & 37.23 $\pm$ 0.01 & 37.27 $\pm$ 0.02 \\ \midrule
\multirow{2}{*}{0.1 < z < 0.2} & $c$ & 37.15 $\pm$ 0.01 & 37.15 $\pm$ 0.01 & 37.15 $\pm$ 0.01 & 37.16 $\pm$ 0.01 \\
 & $c_p$ & 37.15 $\pm$  0.01 & 37.17 $\pm$ 0.01 & 37.17 $\pm$ 0.01 & 37.19 $\pm$ 0.03 \\
\bottomrule
\end{tabular}
\caption{Coefficients of the zero point for the M-Z relation. The coefficients $c$ and $c_p$ correspond to the zero point of the control pair samples, respectively. The rest of the fitted coefficients of the third order polynomial are given in Fig. \ref{GAMAfid}d.}
\label{mzcoeff}
\end{table*}


\section{The effect of pair separation on SFR and Z as a function of redshift} \label{pairsepappendix}

As mentioned in \S 4, in this appendix we show the SFR and Z differences for the redshift range 0$<$z$<$0.1. Due to the redshift cut, the number of galaxies is lower, and therefore some sub-samples are missing. The general tendencies are very similar to the ones showed in section. \ref{pairsec}, although with higher dispersion.

\begin{figure*}
\center
\includegraphics[trim={0 10cm 0 0cm},clip,scale=0.5]{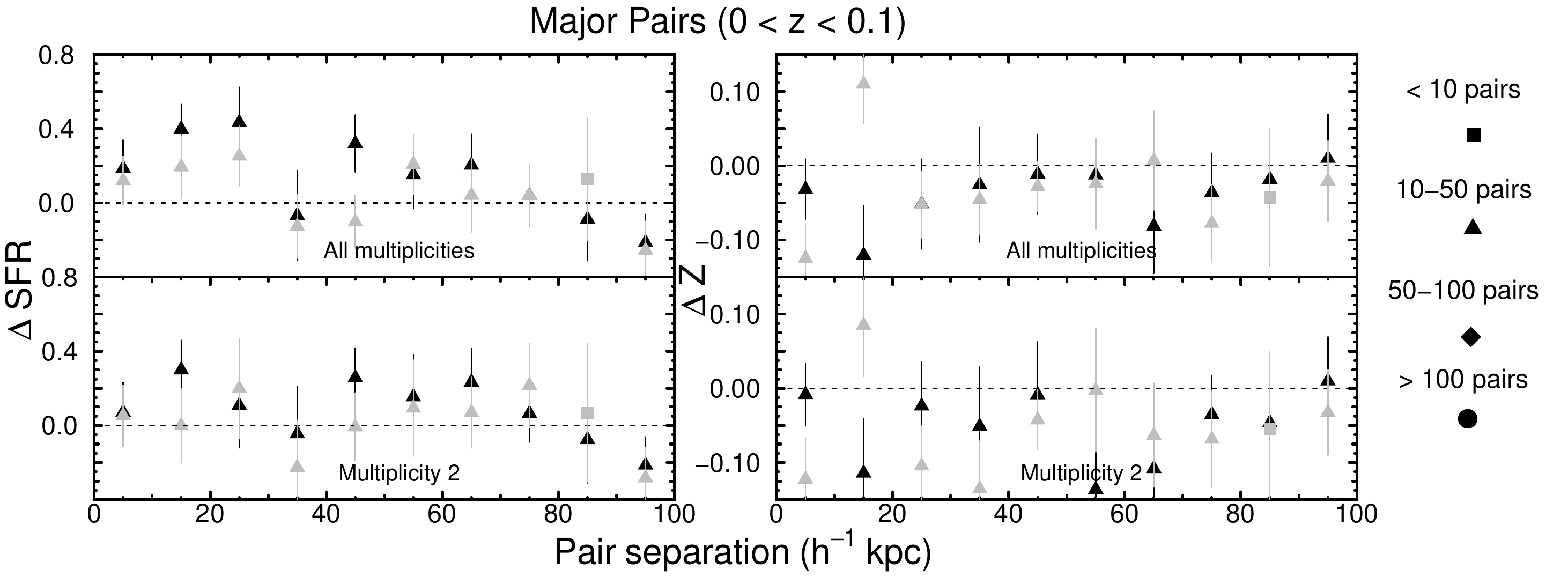}
\caption{Same symbols as in in Fig. \ref{allmajo} for major pairs at 0 $<$ z $<$ 0.1}
\label{projsepz01ma}
\end{figure*}

\begin{figure*}
\center
\includegraphics[trim={0 6cm 0 0cm},clip,scale=0.5]{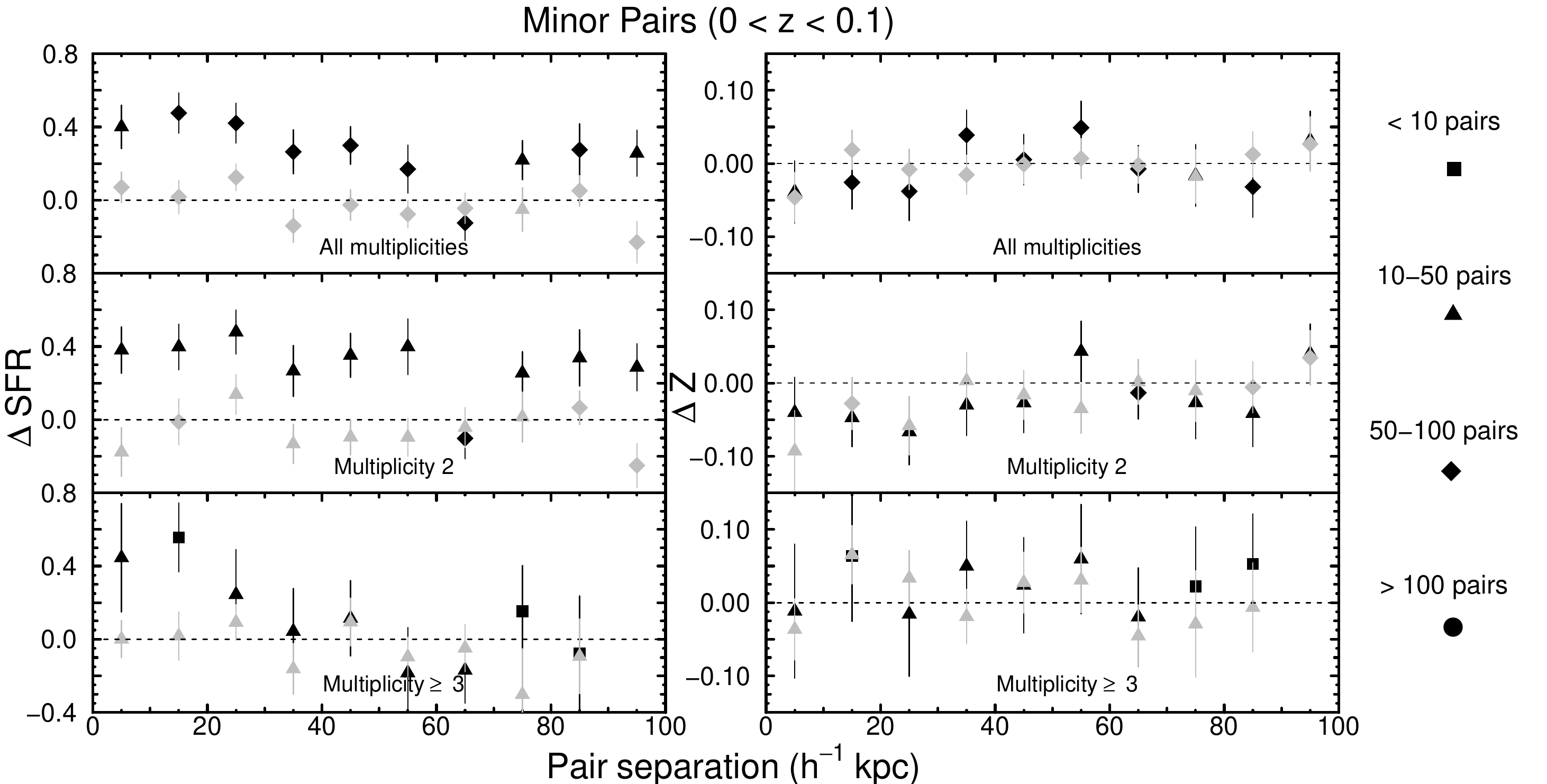}
\caption{Same symbols as in in Fig. \ref{allmajo} for minor pairs at 0 $<$ z $<$ 0.1}
\label{projsepz01min}
\end{figure*}


\bsp	
\label{lastpage}
\end{document}